\documentclass[useAMS,usenatbib,usegraphicx]{mn2e}
\usepackage{times}
\usepackage{amssymb,amsmath,color,soul}
\usepackage{deluxetable}
\usepackage[colorlinks = true,linkcolor = blue,urlcolor  = blue,citecolor = blue,anchorcolor = blue]{hyperref}
\usepackage[caption = false]{subfig}
\usepackage{breqn,color}
\usepackage{float}
\DeclareRobustCommand{\ion}[2]{%
\relax\ifmmode
\ifx\testbx\f@series
{\mathbf{#1\,\mathsc{#2}}}\else
{\mathrm{#1\,\mathsc{#2}}}\fi
\else\textup{#1\,{\mdseries\textsc{#2}}}%
\fi}

\newcommand{\kms}{$\rm~km~s^{-1}$}

\title[Metallicity gradients in local field star-forming galaxies]{Metallicity gradients in local field star-forming galaxies: Insights on inflows, outflows, and the coevolution of gas, stars and metals}

\author[Ho et al.]
{I-Ting Ho$^1$, Rolf-Peter Kudritzki$^{1,2}$, Lisa J. Kewley$^{1,3}$, H. Jabran Zahid$^4$, 
\newauthor Michael A. Dopita$^{3,5}$, Fabio Bresolin$^1$, and David S. N. Rupke$^6$ \\
$^1$Institute for Astronomy, University of Hawaii, 2680 Woodlawn Drive, Honolulu, HI 96822, USA \\
$^2$University Observatory Munich, Scheinerstr. 1, D-81679 Munich, Germany\\
$^3$Research School of Astronomy and Astrophysics, Australian National University, Cotter Road, Weston ACT 2611, Australia \\
$^4$Harvard-Smithsonian Center for Astrophysics, 60 Garden Street MS-20, Cambridge, MA 02138, USA \\
$^5$Astronomy Department, King Abdulaziz University, P.O. Box 80203, Jeddah, Saudi Arabia \\
$^6$Department of Physics, Rhodes College, Memphis, TN 38112, USA }

\begin{document}
\date{Accepted 2015 January 10. Received 2015 January 9; in original form 2014 October 23}

\label{firstpage}
\maketitle

\begin{abstract}

We present metallicity gradients in 49 local field star-forming galaxies. We derive gas-phase oxygen abundances using two widely adopted metallicity calibrations based on the [\ion{O}{iii}]/H$\beta$, [\ion{N}{ii}]/H$\alpha$ and [\ion{N}{ii}]/[\ion{O}{ii}] line ratios. The two derived metallicity gradients are usually in good agreement within $\pm0.14~{\rm dex}~R_{25}^{-1}$ ($R_{25}$ is the {\it B}-band iso-photoal radius), but the metallicity gradients can differ significantly when the ionisation parameters change systematically with radius. We investigate the metallicity gradients as a function of stellar mass ($ 8<{\log}(M_*/M_\odot)<11$) and absolute {\it B}-band luminosity ($ -16 > M_B > -22$). When the metallicity gradients are expressed in $\rm dex~kpc^{-1}$, we show that galaxies with lower mass and luminosity, on average, have steeper metallicity gradients. When the metallicity gradients are expressed in ${\rm dex}~R_{25}^{-1}$, we find no correlation between the metallicity gradients, and stellar mass and luminosity. We provide a local benchmark metallicity gradient of field star-forming galaxies useful for comparison with studies at high redshifts. We investigate the origin of the local benchmark gradient using simple chemical evolution models and observed gas and stellar surface density profiles in nearby field spiral galaxies. Our models suggest that the local benchmark gradient is a direct result of the coevolution of gas and stellar disk under virtually closed-box chemical evolution when the stellar-to-gas mass ratio becomes high ($\gg0.3$). These models imply low current mass accretion rates ($\lesssim0.3\times\rm SFR$), and low mass outflow rates ($\lesssim3\times\rm SFR$) in local field star-forming galaxies.
\end{abstract}


\section{Introduction}

The content of heavy elements in a galaxy is one of the key properties for understanding its formation and evolutionary history. The gas-phase oxygen abundance in the interstellar medium (ISM) of a galaxy (or ``metallicities''), defined as the number ratio of oxygen to hydrogen atom and commonly expressed as $\rm 12+\log(O/H)$, is regulated by various processes during the evolutionary history of a galaxy. While the oxygen is predominately synthesised in high-mass stars ($>8M_\odot$) and subsequently released to the ISM by stellar winds and supernova explosion, the oxygen in the ISM could also be expelled to the circumgalactic medium, or potentially become gravitationally unbound, via feedback processes, e.g. galactic-scale outflows. Gas inflows triggered by mergers and inflows of pristine gas from the intergalactic medium could also dilute the metallicity of a galaxy \citep[e.g.][]{Kewley:2006uq,Rupke:2008fk,Kewley:2010kx,Rupke:2010lr,Rupke:2010fk}. The relations between metallicity and other fundamental properties of galaxies can place tight constraints on the processes governing the evolution of galaxies.

The correlation between global metallicity and stellar mass in star-forming galaxies, i.e. the mass-metallicity relation, is one of the fundamental relations for measuring the chemical evolution of galaxies \citep{Lequeux:1979lr,Tremonti:2004fk}. Whilst the mass-metallicity relation was first established locally, modern spectroscopic surveys have enabled precise measurements of metallicity out to high redshifts on large numbers of galaxies \citep[e.g.,][]{Savaglio:2005qy,Erb:2006fk,Zahid:2011lr,Zahid:2013kx,Zahid:2014fj,Wuyts:2014yq,Steidel:2014rt,Sanders:2014vn} and laid the foundation for subsequent investigation into the physical origin of the relation. Various physical processes including metal enriched outflows \citep[e.g.,][]{Larson:1974lr,Tremonti:2004fk}, accretion of metal-free gas \citep[e.g.][]{Dalcanton:2004lr}, and variation in the initial mass function \citep{Koppen:2007fk} or star formation efficiency \citep[e.g.,][]{Brooks:2007fk,Calura:2009qy} have all been proposed to be responsible for shaping the mass-metallicity relation. The mass-metallicity relation may also have an additional dependency on star formation rate  \citep[SFR; e.g.,][]{Mannucci:2010lr,Lara-Lopez:2010uq,Yates:2012lr}. Spatially resolved studies have shown that the mass-metallicity relation also holds on smaller scales for individual star-forming regions within galaxies \citep{Rosales-Ortega:2012lq}. Recent work suggests that the mass-metallicity relation could be a direct result of some more fundamental relations between metallicity, stellar and gas content \citep{Zahid:2014uq,Ascasibar:2014fk}.

The spatial distribution of metals in a disk galaxy can also provide critical insight into its mass assembly history. Disk galaxies in the local Universe universally  exhibit negative metallicity gradients, i.e., the centre of a galaxy has a higher metallicity than the outskirts \citep[e.g.,][and references therein]{Zaritsky:1994lr,Moustakas:2010fk,Rupke:2010fk,Sanchez:2014fk}. In several cases, where measurements are possible out to very larger radii ($2\times R_{25}$\footnote{Radius of the 25th magnitude/arcsec$^2$ isophote in {\it B}-band.}), the metallicity gradients flatten in the outer disks, suggesting inner-to-outer transportation of metals via mechanisms such as galactic fountains \citep[e.g.,][]{Werk:2011lr,Bresolin:2012lr,Kudritzki:2014fj,Sanchez:2014fk}. Extreme examples of metal mixing occurs in interacting galaxies, where the non axis-symmetric potential induces radial inflows of gas. Both observations and simulations confirm that mergers of disk galaxies present shallower metallicity gradients than isolated disk galaxies due to effective gas mixing \citep[e.g.,][]{Kewley:2010kx,Rupke:2008fk,Rupke:2010lr,Rupke:2010fk,Torrey:2012ai,Rich:2012oq}.

Sophisticated modelling of the evolution of metallicity gradients in disk galaxies has shed light on the formation, gas accretion, and star formation history of the disks. While the details vary from model to model, typical assumptions of inside-out disk growth, no radial matter exchange, and closed-box chemical evolution can successfully reproduce the current gradients in local galaxies \citep[e.g.,][]{Chiappini:2001fk,Fu:2009vn}. However, some models predict that metallicity gradients steepen with time  (e.g., \citealt{Chiappini:1997fk,Chiappini:2001fk}; see also \citealt{Mott:2013mz} who included radial inflow), while others predict the opposite \citep[e.g.,][]{Molla:1997mz,Prantzos:2000vn,Fu:2009vn,Pilkington:2012lr}. Testing the model predictions using observations of high redshift galaxies are challenging \citep[e.g.,][ and references therein]{Yuan:2011qy,Jones:2010uq,Jones:2013kx}. In addition, systematic effects from insufficient resolution and/or binning of the data unfortunately can seriously the reliability of metallicity gradients measured at high redshifts \citep{Yuan:2013gf,Mast:2014lr}.

Statistical studies of metallicity gradients in the local Universe provide an alternative approach to constrain the theoretical simulations. Although the measurements are time-consuming, sample sizes of few tens of galaxies have been achieved in the past using long-slit spectroscopy. These samples gave intriguing (and sometimes contradictory) correlations between metallicity gradients and physical properties of the disk galaxies. For example, barred galaxies tend to exhibit shallower metallicity gradients than non-barred galaxies even when galaxy sizes are taken into account \citep[e.g.,][]{Vila-Costas:1992fj,Zaritsky:1994lr}, but such discrepancy is insignificant in some recent studies \citep{Sanchez:2014fk}. For unbarred galaxies, galaxies with higher {\it B}-band luminosity or higher total mass have shallower metallicity gradients \citep{Vila-Costas:1992fj,Garnett:1997fj}; however, such behaviour is not pronounced in some measurements \citep[e.g.,][]{van-Zee:1998yq,Prantzos:2000vn}. Some studies find that non-barred galaxies show a statistically significant correlation between metallicity gradient and Hubble Type, where earlier types have shallower metallicity gradients than later types \citep{Vila-Costas:1992fj,Oey:1993gf}, but considerable scatter exists in other measurements \citep{Zaritsky:1994lr}. Most studies find no correlation when metallicity gradients are normalised to some scale-length (i.e., $R_{25}$, the disk scale-length $R_d$, or the effective radius $R_e$\footnote{The effective radius is the radius at which the integrated flux is half of the total one. Comparing to the disk scale-length for the classical exponential profile, $R_e = 1.67835 R_d.$}). The contradictory results of some earlier studies might be due to the small sample sizes and inconsistent methodologies of measuring metallicity gradients.  Applying different metallicity diagnostics can introduce considerable systematic errors \citep{Kewley:2008qy}. 

Advances in instrumentation such as multi-slit spectroscopy and wide-field integral field spectroscopy (IFS) is in the process of revolutionising statistical studies of metallicity gradients in the local Universe \citep[e.g.,][]{Sanchez:2014fk}. Large on-going and future large IFS surveys include the Calar Alto Legacy Integral Field Area Survey \citep[CALIFA;][]{Sanchez:2012fj}, the Sydney-AAO Multi-object Integral field spectrograph (SAMI) Survey \citep{Croom:2012qy,Bryant:2014fj,Allen:2014lr}, the Mapping Nearby Galaxies at Apache Point Observatory (MaNGA) Survey, the Hector Survey \citep{Lawrence:2012fk,Bland-Hawthorn:2014lr}, and many others. These IFS surveys are not only extremely efficient in collecting large numbers of spectra simultaneously and seamlessly across an entire galaxy, but also have desirable wavelength coverage to capture multiple key emission lines for deriving metallicity. Such features pose a unique opportunity to eliminate systematic errors using statistical approaches. 

In this paper, we study the metallicity gradients in a sample of 49 local field star-forming galaxies. We derive metallicity gradients using different abundance calibrations and discuss potential systematic effects induced by the calibrations adopted. We investigate whether metallicity gradients in field star-forming galaxies correlate with their physical properties. We show that there is a common metallicity gradient in local field star-forming galaxies and we provide some benchmark values. Finally, we adopt simple chemical evolution models to explain the formation of the common metallicity gradient. 

The paper is structured as follows. We describe our samples, observations and data reduction in Section~\ref{sec-samples}, and our data analysis in Section~\ref{sec-analysis}. In Section~\ref{sec-derive-z-dz-dr}, we detail our methodology of deriving the metallicity, ionisation parameter, and metallicity gradients. In Section~\ref{sec-result}, we present our measurements of metallicity gradients, discuss the systematic effects, and compare the metallicity gradients with stellar masses and absolute {\it B-}band magnitudes. We 
provide a benchmark metallicity gradient in Section~\ref{sec-benchmark-gradient} and investigate the origin of the benchmark gradient in Section~\ref{sec-why-benchmark-gradient} using the simple chemical evolution models. Finally, a summary and conclusions are given in Section~\ref{sec-summary}. Through out this paper, we assume the standard $\Lambda$ cold dark matter cosmology with $ H_0 = 70~\rm km~s^{-1}~Mpc^{-1}$, $\rm \Omega_M = 0.3$ and $\rm \Omega_{\Lambda} = 0.7$.

\section{Samples}\label{sec-samples}

The 49 galaxies studied in this work are drawn from various sources, including literature data, public data and our targeted observations. We describe the four samples in the next four subsections. The focus of this paper is to investigate field star-forming galaxies, and therefore we select only field galaxies that are not undergoing major mergers and not in close pairs; none of our galaxies have massive companions (i.e. stellar mass higher than one-third of the main galaxies) within 70~kpc in projection and $1000~\rm km~s^{-1}$ in line-of-sight velocity separation.

\subsection{CALIFA galaxies in Data Release 1}
We use the publicly available IFS data from the first data release (DR1) of the CALIFA survey. A full description of the survey design, including details of target selection and data reduction scheme, can be found in \citet{Sanchez:2012fj}. Specific details related to the DR1 can be found in \citet[see also \citealt{Walcher:2014qf}]{Husemann:2013yq}. Below we briefly summarise the information relevant to this work. 

The CALIFA DR1 contains reduced IFS data of 100 local galaxies ($0.005<z<0.03$) covered by the Sloan Digital Sky Survey \citep[SDSS;][]{York:2000qe,Abazajian:2009kx}. In this study, we focus only on spiral galaxies that are not undergoing a major merger, and not in close pairs. Extremely edge-on systems with inclination angle larger than $70^\circ$ are excluded from our analysis since de-projecting radial distance is uncertain. Systems without enough sufficiently high signal-to-noise spaxels ($\rm S/N>3$) to measure emission line fluxes are also excluded. In total, 21 CALIFA galaxies are analysed. 

The released CALIFA datacubes, as processed by the CALIFA automatic data reduction pipeline, have a spatial size of $\sim74\arcsec\times64\arcsec$ on a rectangular $1\arcsec$ grid. The point spread function, as measured from field stars in the datacubes, has a median full-width measured at half-maximum (FWHM) of $3.7\arcsec$. Every CALIFA galaxy is observed using the fibre bundle integral field unit (IFU) PPak, and two different setups with the Potsdam Multi-Aperture Spectrophotometer \citep[PMAS;][]{Roth:2005mz,Kelz:2006ly}. The low-resolution (V500) and high-resolution (V1200) setup cover wavelength ranges of $\sim3745\mbox{--}7500$~\AA\ and $\sim3650\mbox{--}4840$~\AA, respectively. The V500 reduced datacubes have a FWHM spectral resolution of 6.0\AA\ ($\rm R \sim850$) and a spectral channel width of 2.0\AA. The V1200 reduced datacubes have a FWHM spectral resolution of 2.3\AA\ ($\rm R \sim1650$) and a spectral channel width of 0.7\AA.

\subsection{WiFeS galaxies}
 
The CALIFA sample was selected based on the angular iso-photal diameter of the galaxies ($45\arcsec<D_{25}<80\arcsec$). Therefore, the CALIFA sample is inevitably biased towards galaxies of higher mass ($\gtrsim10^{9}~M_\odot$). To probe metallicity gradients in galaxies of lower mass in a statistically significant way, we conducted supplemental IFS observations to specifically target low mass systems (i.e., ${\log}(M_*/M_\odot) = 8\mbox{--} 9$). We first selected a mother sample of low-mass galaxies from the SDSS Data Release 7 value-added catalog constructed by the MPA/JHU group\footnote{\url{http://www.mpa-garching.mpg.de/SDSS/DR7/}}. We used the stellar mass derived by the MPA/JHU group as a reference for target selection \citep{Kauffmann:2003yq,Salim:2007ly}; the final stellar masses adopted in this work are derived separately and consistently for all our samples. As a result of this selection, one of the WiFeS galaxies presented has a substantially larger final stellar mass (${\log}(M_*/M_\odot) \sim10.2$) due to the incorrect apertures adopted by MPA/JHU. We remove galaxies with AGN from our mother sample using the optical line ratios [\ion{N}{ii}]/H$\alpha$ and [\ion{O}{iii}]/H$\beta$ \citep{Kewley:2006lr}. We further constrained the mother sample to have low inclination disks and spatial extent comparable to the Wide Field Spectrograph (see Section~\ref{sec-wifes-observation-and-data-reduction}). We also visually confirmed that these galaxies are not undergoing major merger and do not have massive companions. From the mother sample, we then selected our final targets based on observability and instrumental sensitivity. In total, we observed 19 galaxies, 10 of which yield reliable metallicity gradients and are presented in this paper. 

\floatplacement{figure}{H}
\begin{figure}
\centering
\includegraphics[width=8.5cm]{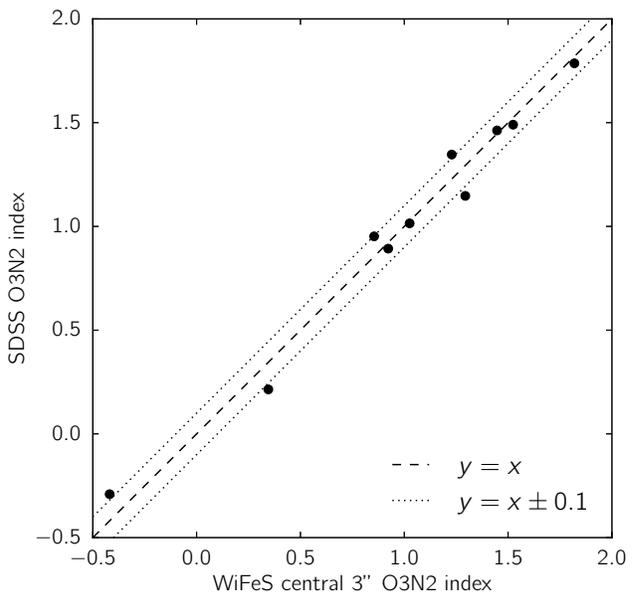} 
\caption{Comparison of the observed O3N2 index (\mbox{Equation~\ref{eq-o3n2}}) between those from the WiFeS and the SDSS data. Each dot corresponds to one of the 10 WiFeS galaxies that were also observed by the SDSS spectroscopic survey. To derive the O3N2 index from the WiFeS data, we extract line fluxes in 3\arcsec apertures at the locations of the SDSS fibres. The O3N2 index derived from the two datasets are consistent within approximately 0.1 dex. }\label{compare_wifes_sdss}
\end{figure}

\subsubsection{Observation and data reduction}\label{sec-wifes-observation-and-data-reduction}
We observed our low-mass galaxies using the WiFeS on the 2.3-m telescope at Siding Spring Observatory in December 2012 and April 2013. WiFeS is a dual beam, image-slicing IFU consisting of 25 slitlets. Each slitlet is $38\arcsec$ long and $1\arcsec$ wide, yielding a $25\arcsec\times38\arcsec$ field of view. For a thorough description of the instrument, see \cite{Dopita:2007kx} and \citet{Dopita:2010yq}. Each galaxy was observed with the blue and red arms simultaneously using the B3000 and R7000 gratings, respectively.  All galaxies were observed with a single WiFeS pointing except for J031752.75-071804 where we adopted a two-point mosaic. The typical exposure time is $\sim1-2$ hours per galaxy under seeing conditions of $1.5-2.5\arcsec$. 

We reduce the data using the custom-built data reduction pipeline {\scshape pywifes} \citep{Childress:2014fr}. The final reduced data consist of two datacubes on $1\arcsec\times1\arcsec$ spatial grids for each galaxy. The blue cube covers $\sim3500\mbox{--}5700$\AA\ with a FWHM velocity resolution of $\rm\sim100~km~s^{-1}$ at H$\beta$ ($\sim1.7$\AA\ or $\rm R\sim3000$) and a spectral channel width of $\sim0.8$\AA. The red cube covers $\sim5500\mbox{--}7000$\AA\ with a FWHM velocity resolution of $\rm\sim40~km~s^{-1}$ at H$\alpha$ ($\sim0.9$\AA\ or $\rm R\sim7000$) and a spectral channel width of $\sim$0.4\AA. 

To compare our reduced WiFeS datacubes with the SDSS fibre spectra, we present in Figure~\ref{compare_wifes_sdss} the emission line ratios, the O3N2 indexes (see below; Equation~\ref{eq-o3n2}), derived from the two datasets. For the SDSS data, we adopt the line fluxes from the MPA/JHU value-added catalog; for the WiFeS data, we extract line fluxes within the corresponding fibre apertures from the emission line maps (see below). Figure~\ref{compare_wifes_sdss} demonstrates that the O3N2 indexes from the WiFeS and SDSS data are consistent within approximately 0.1 dex. 

\floatplacement{figure}{H}
\begin{figure*}
\centering
\includegraphics[width = 17cm]{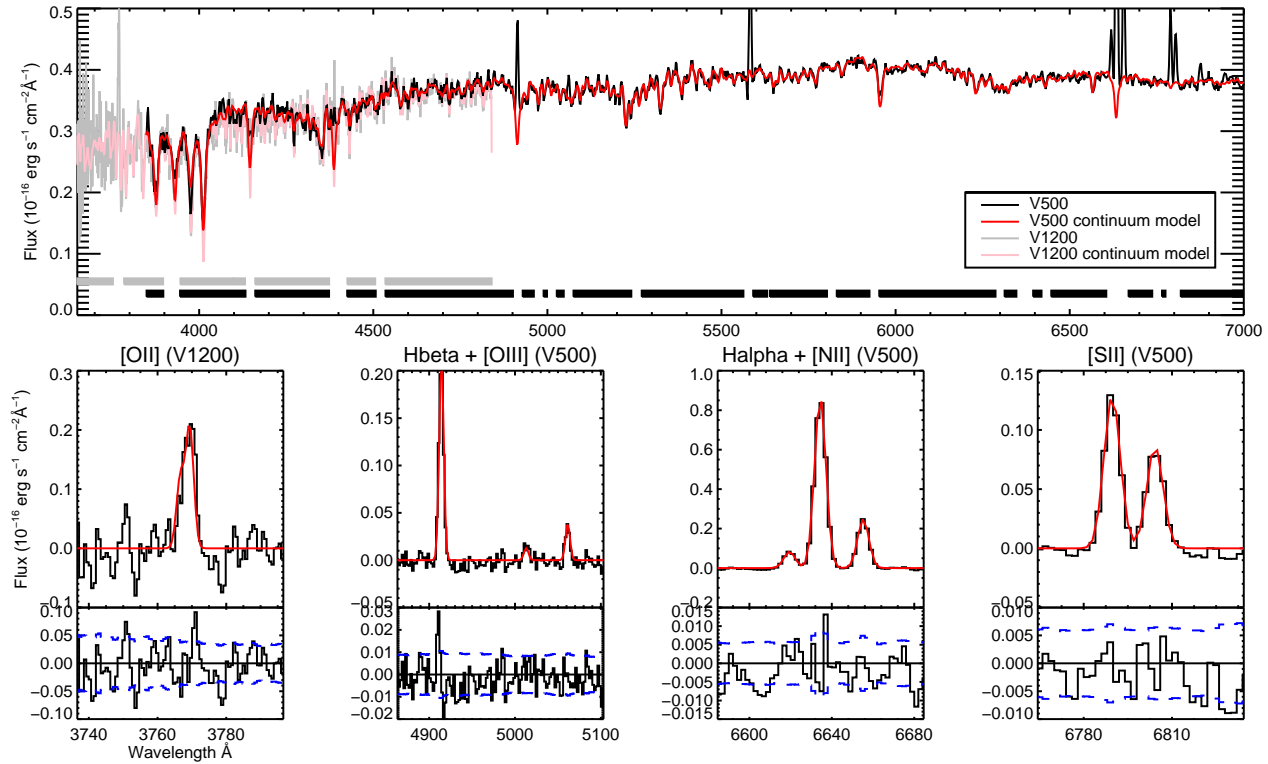}
\caption{An example of the spectral fitting approach applied on the CALIFA data (see Section~\ref{sec-emission-line-maps} for details). The grey and black thick lines in the upper panel indicate the wavelength ranges where the data (black: V500 data; grey: V1200 data) are adopted to constrain the continuum models (red: V500; pink: V1200). Bad channels, the vicinity of strong emission lines and sky lines are excluded from the fit. The four middle panels show emission lines (red) fit to the continuum subtracted spectra (black).  All lines are fit simultaneously and share the same velocity and velocity dispersion. The bottom four panels show the residuals, and the blue dashed lines indicate the $\pm1\sigma$ noise levels. 
}\label{figure2}
\end{figure*}

\subsection{Galaxies from \citet{Sanchez:2012lr}}\label{sec-s12}

To further increase our sample size, we analyse 9 field galaxies presented in \citet[][hereafter S12]{Sanchez:2012lr} with low inclination angles ($<50^\circ$).  
 
S12 studied $\sim2600$ H\textsc{ii} regions in 38 nearby galaxies selected from the PINGS survey \citep{rosales-ortega:2010} and \citet{Marmol-Queralto:2011lr}. All 38 galaxies were observed with PPak and PMAS, which deliver spectral coverages similar to the CALIFA data of $\sim3700-6900$\AA. S12 applied a semi-automatic procedure, {\scshape HiiExplorer}, to search for H\textsc{ii} regions in IFU data under the assumptions that H\textsc{ii} regions are peaky and isolated line-emitting structures with typical physical sizes of few hundred parsecs (more details in S12). A very similar spectral fitting approach (to Section~\ref{sec-emission-line-maps}) was applied on synthetic spectra of H\textsc{ii} regions to decouple the underlying stellar contribution from line emissions. The final public flux catalogs contain seven strong lines, {[\ion{O}{ii}]~$\lambda\lambda$3726,3729}, H$\beta$, {[\ion{O}{iii}]~$\lambda$5007}, {[\ion{O}{i}]~$\lambda$6300}, H$\alpha$, {[\ion{N}{ii}]~$\lambda$6583}, and {[\ion{S}{ii}]~$\lambda\lambda$6716,6731}. 
 
All the 9 galaxies analysed in this study have multiple bright H\textsc{ii} regions measured in all the strong lines including [\ion{O}{ii}]~$\lambda\lambda$3726,3729, which allows reliable constraints simultaneously on metallicity and ionisation parameters with different diagnostics. 

\subsection{Galaxies from \citet{Rupke:2010fk}}

\citet[][hereafter R10]{Rupke:2010fk} measured metallicity gradients in interacting and non-interacting galaxies. They show that, on average, interacting systems present shallower metallicity gradients than non-interacting systems. Their control sample comprises 11 non-interacting local galaxies and they measure their metallicity gradients using published emission line data from H\textsc{ii} regions. Two of these control sample galaxies overlaps with the S12 sample. We include their measurements for the remaining 9 galaxies in our analysis. Their methods of correcting for extinction and deriving metallicity are exactly the same as our work and therefore we include their measurements of metallicity gradients without a re-analysis of the data.

\floatplacement{figure}{H}
\begin{figure*}
\includegraphics[width = 17cm]{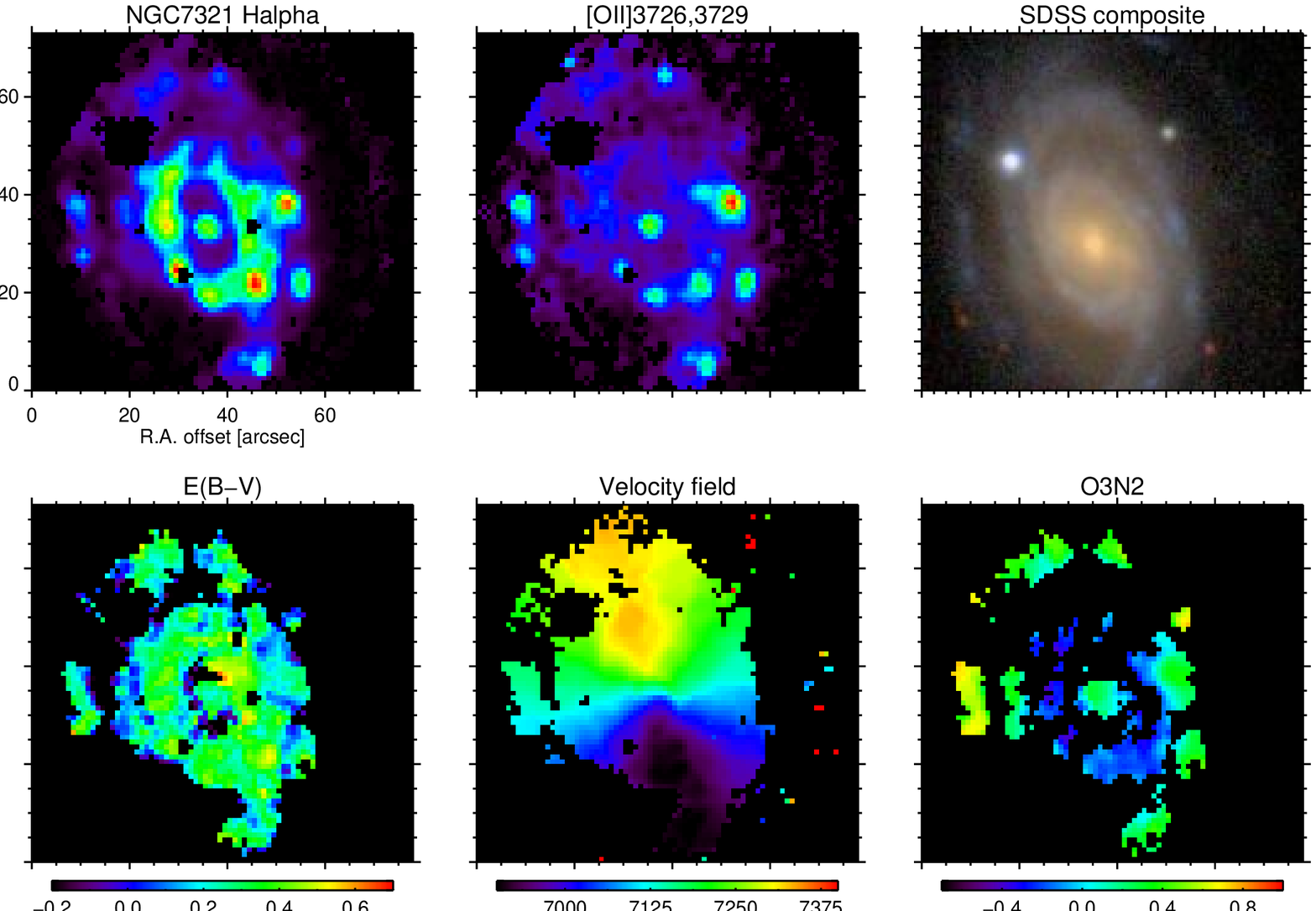}
\vspace{+0.5cm}
\caption{An example of 2D maps from our spectral analysis described in Section~\ref{sec-emission-line-maps}. The first row shows the H$\alpha$ map, [\ion{O}{ii}]~$\lambda\lambda$3726,3729 map, and SDSS 3-colour image of NGC7321, one of the CALIFA galaxies. The second row shows the E(B-V) , velocity field, and O3N2 maps. The bright foreground star in the SDSS image is masked out in all the other maps. }\label{figure3}
\end{figure*}

\section{Analysis}\label{sec-analysis}

\subsection{Emission line maps}\label{sec-emission-line-maps}

To place constraints on metallicity and ionisation parameter using emission line diagnostics, we measure emission line fluxes in each spaxel of CALIFA and WiFeS galaxies by spectral fitting. We use an earlier version of the spectral fitting tool {\scshape lzifu} described in Ho et al. (in preparation; see also \citealt{Ho:2014uq}). The fitting approaches for the two samples are very similar, though some details are fine tuned to accommodate the differences in spectral coverage and resolution between the two datasets. Below, we first elaborate our method for the CALIFA galaxies before describing the different treatments for the WiFeS galaxies. 

Prior to fitting the CALIFA galaxies, we first correct for the known spatial misalignment between the V500 and V1200 datacubes \citep{Husemann:2013yq}. For each CALIFA galaxy, we realign the two cubes by correlating continuum images constructed separately from the two cubes using a wavelength range covered by both settings (4240--4620\AA). Typical misalignments are $\sim1\arcsec$ to $2\arcsec$ while in several extreme cases $\sim3\arcsec$ to $5\arcsec$. We then rescale the V1200 data to match the V500 data using scale factors determined from the median flux density ratios between an overlapped spectral coverage of 4000\AA\ and 4500\AA. Slight differences in flux levels may be a result of imperfect calibration and any residual systematic errors. 

We then perform simple stellar population (SSP) synthesis to remove the underlying stellar continuum before fitting emission lines. To model the stellar continuum, we adopt the penalised pixel-fitting routine \citep[{\scshape ppxf};][]{Cappellari:2004uq} and employ the empirical MIUSCAT\footnote{\url{http://miles.iac.es/pages/ssp-models.php}} SSP models of 13 ages\footnote{0.06, 0.10, 0.16, 0.25, 0.40, 0.63, 1.00, 1.58, 2.51, 3.98, 6.31, 10.00, and 15.85 Gyr} and 4 metallicities\footnote{[M/H] $=$ -0.71, -0.40, 0.00, and 0.22} \citep{Vazdekis:2010lr}  while assuming a Salpeter initial mass function \citep[IMF;][]{Salpeter:1955kx}. Bad pixels, sky lines and the vicinity ($\pm15$\AA) of emission lines are masked prior to fitting the continuum models. After subtracting the continuum models from the data, we further remove low frequency fluctuations in the residuals by fitting fourth order b-spline models. We note that the major goal of performing SSP synthesis fitting is to correct for stellar Balmer absorption; we do not derive stellar age and metallicity from our SSP fits. 

Subsequent to removing the continuum in the V500 and V1200 datacubes separately, we model the strong emission lines ({[\ion{O}{ii}]~$\lambda\lambda$3726,3729}, H$\beta$, {[\ion{O}{iii}]~$\lambda\lambda$4959,5007}, {[\ion{N}{ii}]~$\lambda\lambda$6548,6583}, H$\alpha$, and {[\ion{S}{ii}]~$\lambda\lambda$6716,6731}) as simple gaussians. We perform a bounded value nonlinear least-squares fit using the Levenberg-Marquardt method implemented in IDL \citep[][{\scshape mpfit}]{Markwardt:2009lr}. We constrain (1) all lines to have the same velocity and velocity dispersion, (2) the ratios {[\ion{N}{ii}]~$\lambda$6583}/{[\ion{N}{ii}]~$\lambda$6548} and {[\ion{O}{iii}]~$\lambda$5007}/{[\ion{O}{iii}]~$\lambda$4959} to their theoretical values given by quantum mechanics, (3) the velocity to be between $+600$ \kms\ and $-600$ \kms\ to the systemic velocities as measured from SDSS, and (4) the velocity dispersion to be between 50 and 1000 \kms. 

Figure~\ref{figure2} shows a typical spectrum and spectral fit. This figure best illustrates the above procedure of decoupling stellar contribution in each spaxel from emission lines originated predominantly from H\textsc{ii} regions. Figure~\ref{figure3} shows two emission line maps, SDSS composite image, extinction map, velocity field map, and a key diagnostics line-ratio map of the CALIFA galaxy NGC7321 to demonstrate the final products from our analysis described above.

For fitting the WiFeS galaxies, the above procedure is adopted with some minor modifications to accommodate the different spectral coverages and resolutions.  No correction for misalignment is required for the WiFeS datacubes since the blue and red data were observed simultaneously. SSP synthesis fitting is performed simultaneously on both cubes to take full advantage of the 4000\AA\ break, an important age indicator, captured only in the blue data. Fitting the red side separately would in principle increase the degeneracies in SSP synthesis fitting. We first down-grade the red data to the same spectral resolution as the blue data ($\rm R\sim3000$), mask out noisy parts of the spectra due to poor CCD sensitivities, and merge the two datacubes to form a master datacube which covers $\sim3700\mbox{--}6950$\AA. We then use {\scshape ppxf} and theoretical SSP models, assuming Padova isochrones, of 18 ages\footnote{0.004, 0.006, 0.008, 0.011, 0.016, 0.022, 0.032, 0.045, 0.063, 0.089, 0.126, 0.178, 0.251, 0.355, 0.501, 0.708, 1.000, and 1.413 Gyr.} and 3 metallicities\footnote{$ Z = 0.004$, 0.008, and 0.019.} from \citet{Gonzalez-Delgado:2005lr} to determine the contribution from different stellar populations and the degree of dust extinction. The results are then used to reconstruct continua of the blue and red data at their native spectral resolution. The same line fitting approach is then applied to the continuum subtracted cubes and yields emission line maps.

\subsection{Extinction correction}\label{sec-extinction-correction}
We use a consistent method for all the galaxies to correct the wavelength-dependent extinction caused by dust attenuation. For a given measurement (i.e., a spaxel for the CALIFA and WiFeS samples or an H\textsc{ii} region for the S12 samples), we assume the classical extinction law by \citet{Cardelli:1989qy} with ${\rm Rv = 3.1}$ and ${\rm H}\alpha/{\rm H}\beta=2.86$ under the case-B recombination of $\rm T_e = 10,000~K$ and $\rm n_e = 100~cm^{-3}$ \citep{Osterbrock:2006lq}. This prescription is consistent with that adopted in R10.

\subsection{Other physical quantities}

\subsubsection{Stellar mass $(M_*)$}
We use the {\scshape Le Phare}\footnote{\url{http://www.cfht.hawaii.edu/~arnouts/LEPHARE/lephare.html}} code developed by Arnouts, S. \& Ilbert, O. to estimate the galactic stellar mass. {\scshape Le Phare} compares photometry measurements with stellar population synthesis models, based on a $\chi^2$ template-fitting procedure, to determine mass-to-light ratios, which are then used to estimate the stellar mass of galaxies. The stellar templates of \citet{Bruzual:2003qy} and a Chabrier IMF \citep{Chabrier:2003uq} are used to synthesise magnitudes. The 27 models span three metallicities and seven exponentially decreasing star formation models (${\rm SFR}\propto e^{-t/\tau}$) with $\tau = 0.1, 0.3, 1, 2, 3, 5, 10, 15$ and 30 Gyr. We apply the dust attenuation law from \citet{Calzetti:2000qy} allowing $E(B-V)$ to vary from 0 to 0.6 and stellar population ages ranging from 0 to 13 Gyr. 

Photometric measurements are collected from various sources. For all the 21 CALIFA galaxies, 10 WiFeS galaxies, and 5/9 S12 galaxies, we adopt values from the SDSS DR7 photometry catalog \citep{Abazajian:2009kx} and 2MASS extended source catalog \citep{Skrutskie:2006yq}. The Petrosian magnitudes of SDSS {\it u, g, r, i}, and {\it z}-band are corrected for the foreground Galactic-extinction \citep{Schlegel:1998ys}. The 2MASS {\it J, H},  and $K_s$ magnitudes measured from fit extrapolation are adopted to approximate the total magnitudes. All these galaxies have the 5-band SDSS photometry and the majority (28/36) also have the 3-band 2MASS photometry. For the rest of the four S12 galaxies and the nine R10 galaxies, we collect available {\it U, B, V}, and the 3-band 2MASS photometric measurements from NASA/IPAC Extragalactic Database (NED). All these galaxies have the 3-band 2MASS photometry. The {\it B} and {\it V}-band photometry are available for all galaxies, and the {\it U}-band photometry is available for about half of these galaxies (6/13). These optical photometric measurements are also corrected foreground Galactic-extinction. When reasonable uncertainties of the measurements are unavailable, we assume 0.1~dex for the optical bands and 0.05~dex for the infrared bands. 

We note that the uncertainties in stellar mass are typically dominated by systematic errors. Different stellar mass estimators employ different algorithms and stellar libraries, but the estimated $M_*$ typically agrees within $\rm \sim0.3~dex$, marking the degree of systematic errors in the measurements \citep[e.g.,][]{Drory:2004vn,Conroy:2009rt}.

\subsubsection{Inclination angle}

Inclination angles of CALIFA galaxies are estimated using the conversion provided by \citet{Padilla:2008fk}. The effects of dust extinction and reddening were taken into account in their analysis. An estimate of inclination angle for spiral galaxy is inferred from the measured axis ratio ($b/a$) and $r$-band absolute magnitudes. Axis ratios estimated by the CALIFA team from SDSS $r$-band images are adopted. Inclination angles of WiFeS galaxies are drawn from Hyperleda \citep{Paturel:2003fr}, which assumes the classical Hubble formula \citep{Hubble:1926uq}. Inclination angles of S12 galaxies are taken directly from table 1 in S12, which also refers to values from Hyperleda. Inclination angles of R10 galaxies are taken directly from table 2 in R10, which is a compilation from various references.

\subsubsection{Size and distance}
We compile the sizes and distances of our samples from NED and Hyperleda. $R_{25}$ from Hyperleda is adopted throughout the paper to quantify sizes of the galaxies. Redshift independent distances are available for all R10 galaxies in Table~2 of R10. Redshift independent distances for most S12 galaxies (7/10) are adopted from NED. These redshift independent distances ($\rm d\lesssim30~Mpc$) are measured from the Tully-Fisher relation, tip of the red-giant brach method, planetary nebulae, type-II supernovae, or Cepheid variables. For CALIFA, WiFeS, and the rest of the three S12 galaxies, we adopt Hubble distances inferred from their redshifts.

\section{Derivation of metallicities and metallicity gradients}\label{sec-derive-z-dz-dr}

In this section, we describe our methodology for using emission line ratios to derive metallicities, metallicity gradients and ionisation parameters. 

\subsection{Metallicity}\label{sec-metallicity}

The most direct way to determine an ISM metallicity is by first measuring electron temperatures ($T_e$) with temperature sensitive line ratios, e.g., {[\ion{O}{iii}]~$\lambda$4363} to {[\ion{O}{iii}]~$\lambda$5007}, and then convert emission measures to metallicity after correcting for unseen stages of ionisation. Since {[\ion{O}{iii}]~$\lambda$4363} is typically unavailable or only detected in very limited (i.e. the hottest) regions in IFU surveys, measuring metallicity usually relies on empirical or theoretical calibrations (or a combination of both) based on strong emission lines, such as those available in our samples. Various such calibrations  are available in the literatures and are widely adopted to derive metallicity \citep[e.g.,][]{Kewley:2002fj,Pettini:2004lr}. 

We derive metallicities with two different calibrations elaborated below. Throughout the paper, we express the metallicity is in terms of the number ratio as $\rm 12 + \log(O/H)$.

\subsubsection{ O3N2 index / PP04}
The O3N2 index, defined as 
\begin{equation}\label{eq-o3n2}
{\rm O3N2}\equiv \log { \mbox{[\ion{O}{iii}]~$\lambda$5007}/{\rm H}\beta \over \mbox{[\ion{N}{ii}]~$\lambda$6583}/{\rm H}\alpha },
\end{equation}
is a widely used metallicity diagnostic in the literature. An empirical calibration is provided by \citet[][hereafter PP04]{Pettini:2004lr}. The popularity of this diagnostics arises for two reasons. Firstly, the four lines involved are usually easily measured in local galaxies out to large radii using modern instruments. Secondly, because of the minimum wavelength differences between the two pairs of lines, the O3N2 index is virtually free from systematics caused by the assumption of an extinction law or reddening uncertainties. Nevertheless, the calibration provided by \citet{Pettini:2004lr} is calibrated empirically with a linear fit to metallicities of 137 extragalactic H\textsc{ii} regions (131 with $T_e$-based metallicity and 6 with detailed photoionisation models). Variation of ionisation parameter, $q$, is not considered in the calibration. Neglect of this parameter may cause serious systematic errors in the results. An updated calibration based on many more $T_e$-based metallicities of H\textsc{ii} regions (309) is provided by \citet{Marino:2013vn}. This new calibration presents a significantly shallower slope between the O3N2 index and metallicity than the PP04 calibration. 


\subsubsection{N2O2 index / KD02}
The N2O2 index, defined as
\begin{equation}\label{eq-n2o2}
\rm N2O2 \equiv \log { \mbox{[\ion{N}{ii}]~$\lambda$6583} \over \mbox{[\ion{O}{ii}]~$\lambda\lambda$3726,3729}},
\end{equation}
is an alternative metallicity diagnostic that has been calibrated by \citet[][hereafter KD02]{Kewley:2002fj} using theoretical photoionisation models. We adopt the parametrisation for $q = \rm 2\times10^7 cm~s^{-1}$ in KD02 to derive metallicity. The N2O2 index is insensitive to variation of ionisation parameter by virtue of the similar ionising potential of $\rm N^+$ and $\rm O^+$. Despite the insensitivity to ionisation parameter, N2O2 is not often used in local studies  primarily because some spectrographs are not sensitive enough at $\sim3700$\AA\ to observe [\ion{O}{ii}]~$\lambda\lambda$3726,3729. The N2O2 calibration depends more on the assumed extinction law and extinction estimate than O3N2 due to the larger wavelength separation between [\ion{N}{ii}]~$\lambda$6583 and [\ion{O}{ii}]~$\lambda\lambda$3726,3729. We note that the variation in the N/O ratio with O/H could be the largest uncertainty affecting these strong line diagnostics \citep{Henry:2000lr}. The strength of the [\ion{O}{ii}]~$\lambda\lambda$3726,3729 lines are strongly affected by the electron temperature, governed predominately by the O/H ratio and ionisation parameter, while [\ion{O}{iii}]~$\lambda$5007 is mostly sensitive to the ionisation parameter.

A long standing problem in chemical studies has been that different metallicity calibrations do not return consistent metallicity measurements. \citet{Kewley:2008qy} applied 10 different metallicity calibrations on the same SDSS dataset and found that different calibrations yield different mass-metallicity relations. Both the slopes and the intercepts of the mass-metallicity relation are significantly different from calibration to calibration. By allowing the mass-metallicity relations derived with different calibrations to be converted to the same bases, \citet{Kewley:2008qy} derived empirical conversions between different calibrations. In this paper, metallicities derived using the O3N2 method are subsequently converted to the KD02 scale using the conversion by \citet{Kewley:2008qy}. 

We note that, when deriving metallicities using a certain diagnostic, we only use data with $\rm S/N>3$ on all the lines associated to that particular diagnostic. The same rule also applies to the derivation of the ionisation parameter (see Section~\ref{sec-ionisation-parameter}).

\subsection{The importance of non-thermal excitation and diffuse ionised gas}\label{sec-dig}
It is important to understand that all the above metallicity diagnostics are calibrated empirically or theoretically using H\textsc{ii} regions. This means, by applying the diagnostics, one implicitly assumes that that all the nebular emission originates from photoionisation and heating caused by the extreme ultraviolet photons emitted by O and B-type stars. If other ionisation sources are present, metallicity measurements would be contaminated. Other ionisation sources such as active galactic nuclei (AGNs) can have localised or even global effects on line ratios. Interstellar shocks originating from AGN outflows, supernovae, or stellar winds are also sources of non-thermal radiation and can affect  emission line ratios. To remove measurements affected by non-thermal radiation from our subsequent analyses, we use line ratio diagnostics commonly adopted to distinguish normal from active galaxies \citep[i.e., {[\ion{O}{iii}]~$\lambda$5007/H$\beta$ v.s. [\ion{N}{ii}]~$\lambda$6583/H$\alpha$} or BPT diagram; ][]{Baldwin:1981lr,Veilleux:1987qy}. We adopted the empirically separation line derived from SDSS by \citet[][hereafter K03; see also \citealt{Kewley:2001lr} and \citealt{Kewley:2006lr}]{Kauffmann:2003vn}  to exclude data contaminated by non-thermal excitation. 

Emission from the diffuse ionise gas (DIG; also known as the warm ionised medium) can be a non-negligible component in IFU measurements. The DIG is hot ($\sim10^4$~K) and tenuous gas ($\rm\sim10^{-1}~cm^{-3}$) permeating the interstellar space and extending more than 1~kpc above the disk plane (see \citealt{Mathis:2000fj} and \citealt{Haffner:2009fr} for reviews). The DIG generally presents much lower H$\alpha$ surface brightness than typical H\textsc{ii} regions and displays distinctly different line ratios from H\textsc{ii} regions. The primary excitation sources are thought to be predominately ionising photons escaping and traveling kilo-parcsec distances from O and B-type stars. Measurements in an IFU spaxel can contain both emission from underlying H\textsc{ii} regions and from the DIG along the line-of-sight.

\floatplacement{figure}{!t}
\begin{figure}
\hspace{-0.5cm}
\includegraphics[width=9cm]{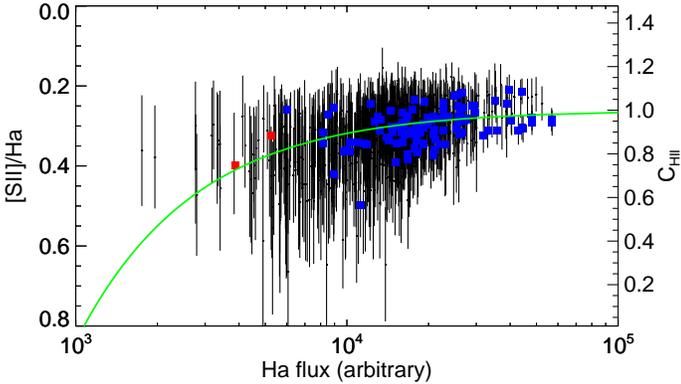}
\caption{An Example of determining spaxels heavily contaminated by the diffuse ionised gas. Details are described in Section~\ref{sec-dig}. The black points are all star-forming spaxels with $>3\sigma$ detections on the [\ion{S}{ii}]~$\lambda\lambda$6716,6731 and H$\alpha$ lines. The blue and red squares are those spaxels with $>3\sigma$ detections on all lines associated with the O3N2 index, i.e., [\ion{O}{III}]~$\lambda$5007, [\ion{N}{ii}]~$\lambda$6583, H$\alpha$, and H$\beta$. The green curve indicates the best fit to the black points using equations~\ref{dig_eq1} and \ref{dig_eq2}, which we adopt to determine the critical H$\alpha$ flux above which the covering fraction of H\textsc{ii} region exceeds 80\%. Data below the critical H$\alpha$ flux (red) are excluded from deriving O3N2 metallicity. }\label{dig_example}
\end{figure}

To quantify the fractional contributions from the DIG and H\textsc{ii} regions in a given spaxel, we adopt a similar approach as \citet{Blanc:2009uq}. Spaxels dominated by the DIG are subsequently removed from our analyses. 
We constrain the contribution from the DIG with the observed the observed [\ion{S}{ii}]/H$\alpha$ ratio
\begin{equation}\label{dig_eq1}
\rm{\mbox{[\ion{S}{ii}]}\over H{\mathnormal\alpha}} \ = \ {\mathnormal Z'} \left[ {\mathnormal C}_{H_{II}} ({\mbox{[\ion{S}{ii}]}\over H{\mathnormal\alpha}})_{\rm H_{II}} + {\mathnormal C}_{DIG} ({\mbox{[\ion{S}{ii}]}\over H{\mathnormal\alpha}})_{DIG} \right], 
\end{equation}
where [\ion{S}{ii}] denotes the total flux of [\ion{S}{ii}]~$\lambda$6716 and [\ion{S}{ii}]~$\lambda$6731. The terms $C_{\rm  H_{II}}$ and $ C_{\rm DIG}$ represent fractions of emission lines originated from H\textsc{ii} regions and the DIG, respectively. The sum of $C_{\rm H_{II}}$ and $C_{\rm DIG}$ is unity. $Z'$ denotes metallicity of the galaxy normalised to that of the Milky Way, i.e. $Z' = Z/Z_{MW}$. Following \citet{Blanc:2009uq}, we adopt the value of $\rm(\mbox{[\ion{S}{ii}]}/H{\mathnormal\alpha})_{\rm H_{II}}$ as 0.11 and $\rm(\mbox{[\ion{S}{ii}]}/H{\mathnormal\alpha})_{DIG}$ as 0.34. These values are supported by observations of H\textsc{ii} regions and the DIG in our Milky Way \citep{Madsen:2006fj}.

Figure~\ref{dig_example} shows a typical $\rm\mbox{[\ion{S}{ii}]}/H{\mathnormal\alpha}$ versus H$\alpha$ flux plot of the CALIFA galaxy NGC6497. All the data points (spaxels) are significantly detected ($>3\sigma$) in [\ion{S}{ii}] and H$\alpha$. Spaxels with high H$\alpha$ fluxes have low (high) $\rm\mbox{[\ion{S}{ii}]}/H{\mathnormal\alpha}$, consistent with the low (high) line ratios of H\textsc{ii} regions (DIG). H\textsc{ii} regions are generally located on spiral arms and DIG is generally located in inter-arm regions.

Similar to \citet{Blanc:2009uq}, we model the covering fractions of H\textsc{ii} regions in each galaxy as a simple function of
\begin{equation}\label{dig_eq2}
C{\rm _{\rm H_{II}}} \ = \  1 - {f_0 \over f({\rm H}\alpha)},
\end{equation}
where $f({\rm H}\alpha)$ is the H$\alpha$ flux.  Combining equation~\ref{dig_eq1} and equation~\ref{dig_eq2}, we fit simultaneously $\rm Z'$ and $f_0$, and show the best fit as the green curve in Figure~\ref{dig_example}. We typically find equation~\ref{dig_eq2} describes the data reasonably well, but the theoretical basis of this functional form is unclear. The best fit provides a guide for imposing a criterion on the H$\alpha$ flux to reject spaxels below a characteristic covering fraction of H\textsc{ii} regions. We exclude all spaxels below an H$\alpha$ flux value at which the corresponding $C_{\rm H_{II}}$ is 0.8 on the best fit curve. Although this criterion seems arbitrarily strict, changing the characteristic covering fraction to zero has only a minor impact on the metallicity gradients of individual galaxies, and none of our conclusions change. 

The reason that rejecting DIG dominated spaxels has limited effect is that the covering fraction cut does not remove many spaxels that have not already been rejected by the S/N cuts for metallicity diagnostics and the BPT criterion (K03) for non-thermal excitation. The strictest criteria in rejecting the data are the S/N cuts on the weak lines such as H$\beta$ and [\ion{N}{ii}]~$\lambda\lambda$6548,6583. Since the DIG has intrinsically low surface brightness, with the current depth in CALIFA and WiFeS samples, most spaxels satisfying multiple $\rm S/N>3$ criteria in line emissions are those with low DIG covering fractions. We demonstrate this in Figure~\ref{dig_example}. The blue and red points are the spaxels satisfying the $\rm S/N>3$ criteria in the O3N2 diagnostic and the criterion on the BPT (K03). The red points are those further rejected by the DIG criterion.

We note that rejecting data dominated by the DIG is only required for spaxel-to-spaxel analysis, i.e. the CALIFA and WiFeS samples. For the S12 galaxies, the fluxes were measured on extracted spectra of H\textsc{ii} regions (i.e., binning spaxels in the vicinity of bright and compact H$\alpha$ knots). For R10, the flux measurements were performed by placing long-slits on H\textsc{ii} regions. DIG contamination is expected to be negligible in these two cases.

In principle, after the fractional contribution is determined, one should be able to subtract the contribution from the DIG and derive the fluxes from underlying H\textsc{ii} regions (as in \citealt{Blanc:2009uq}). However, line ratios involving [\ion{O}{ii}]~$\lambda\lambda$3726,3729 for the DIG are not well constrained and could produce large scatter \citep[{e.g., [\ion{O}{ii}]~$\lambda\lambda$3726,3729/H$\alpha$};][]{Mierkiewicz:2006kx}. Even for the well-measured ratios, such as [\ion{S}{ii}]/H$\alpha$ and [\ion{N}{ii}]/H$\alpha$, considerable scattering and correlation between [\ion{S}{ii}]/[\ion{N}{ii}] could complicate the corrections and induce potential systematic errors \citep[][]{Madsen:2006fj}. In this work, we adopt the simplest approach of rejecting inappropriate data. Comprehensive theoretical modelling and observational constraints of DIG are crucial for extracting more information from deeper IFU data.

\floatplacement{figure}{H}
\begin{figure*}
\centering
\includegraphics[width=16.5cm]{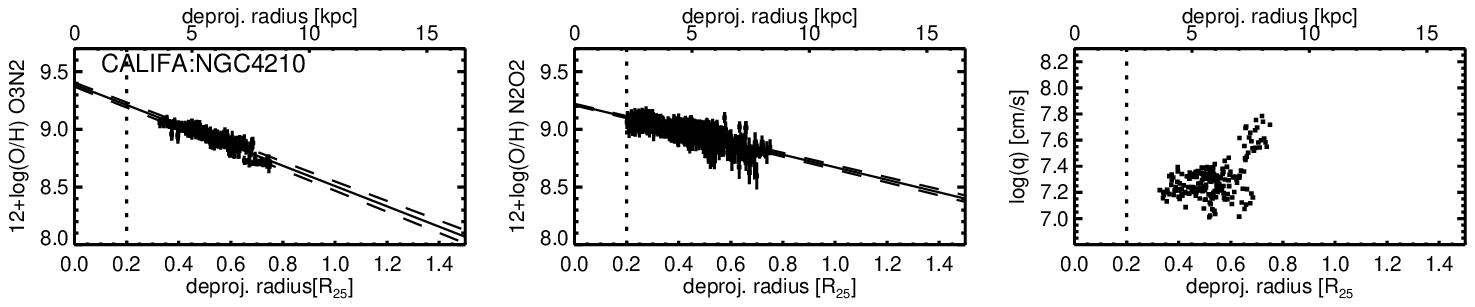}
\includegraphics[width=16.5cm]{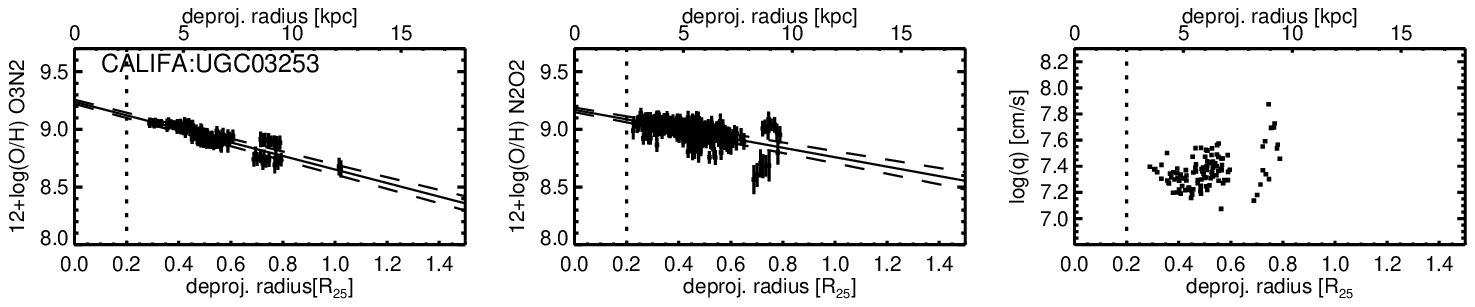}
\caption{{\it Left and middle panels:}  Metallicity gradients of individual CALIFA galaxies measured using the two different abundance diagnostics (see Section~\ref{sec-metallicity}). The straight lines indicate the best fits, and the dashed lines indicate $\pm1\sigma$ errors. The errors of the intercepts and the slopes are estimated from bootstrapping (see Section~\ref{sec-measuring-metallicity-gradients}). {\it Right panels:} ionisation parameter as a function of radius. The ionisation parameters are derived using the [\ion{O}{iii}]/[\ion{O}{ii}] diagnostic (KK04; see Section~\ref{sec-ionisation-parameter}). Each point in these plots corresponds to one IFU spaxel with significant ($>3\sigma$) detections on all the emission lines associated with the diagnostics. Spaxels contaminated by non-thermal excitation or dominated by DIG emission are rejected (see Section~\ref{sec-dig}). The vertical dashed lines correspond to the radial cutoff within which the data are not considered in constraining the disk metallicity gradients. }\label{califa_gradient}
\end{figure*}
\floatplacement{figure}{H}
\begin{figure*}
\centering
\includegraphics[width=16.5cm]{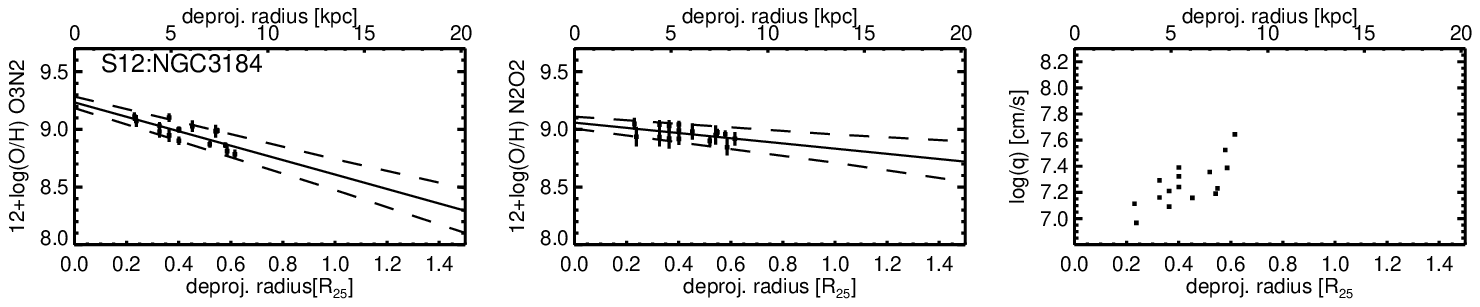} 
\includegraphics[width=16.5cm]{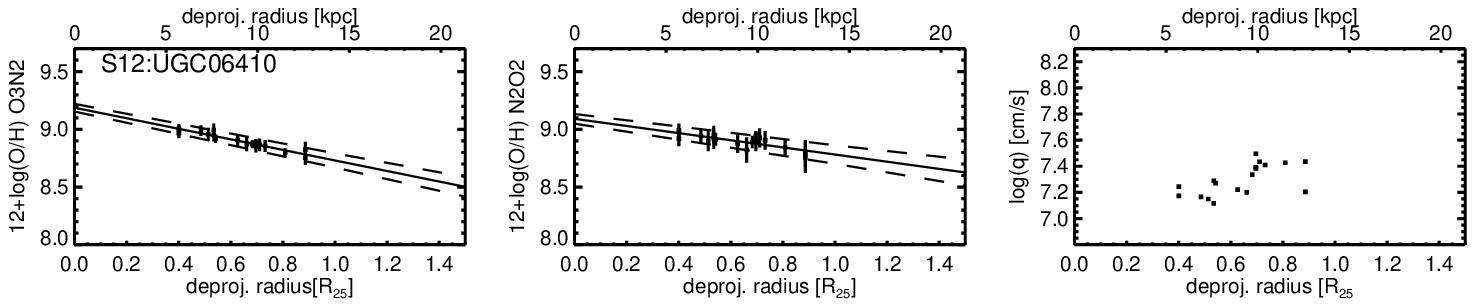} 
\caption{Same as Figure~\ref{califa_gradient}, but for the S12 galaxies. Each point corresponds to one H\textsc{ii} region extracted from the IFU data (see Section~\ref{sec-s12} and S12 for details). }\label{s12_gradient}
\end{figure*}

\subsection{Ionisation parameter}\label{sec-ionisation-parameter}

The ionisation parameter is quantified as the ionising photon flux through a unit area divided by the local number density of hydrogen atoms. The ionisation parameter can be measured by taking the ratio of high ionisation to low ionisation species of the same atom. Using the strong lines available in this study, we measure the ionisation parameter using [\ion{O}{iii}]~$\lambda$5007/[\ion{O}{ii}]~$\lambda\lambda$3726,3729 (hereafter, [\ion{O}{iii}]/[\ion{O}{ii}]). A theoretical calibration was first presented in KD02 and a more user-friendly parametrisation is given in \citet[][hereafter KK04]{Kobulnicky:2004lr}. The latter is adopted in this paper. As emphasised in KD02, in addition to ionisation parameter, the [\ion{O}{iii}]/[\ion{O}{ii}] ratio also strongly depends on metallicity that needs to be known a priori. We adopt the metallicity from the N2O2 calibration to derive the ionisation parameter for each spectrum.

\subsection{Measuring metallicity gradients}\label{sec-measuring-metallicity-gradients}

To derive metallicity gradients, we first convert the flux ratios to metallicities. For bulge-dominated galaxies or those with obvious bar structures in the CALIFA sample, we do not include data within $(0.1\mbox{--} 0.2)\times R_{25}$. At the centres of these generally high-mass systems, the optical spectrum is dominated by the stellar continuum with strong Balmer absorption originating from an old stellar population. Correcting for Balmer absorptions in these regions is less robust and more sensitive to both the SSP models and the algorithms adopted \citep[e.g.,][ and references therein]{Cid-Fernandes:2014qy}. The WiFeS galaxies do not suffer from this contamination because these low-mass systems are typically emission dominated even at the centres. The S12 and R10 galaxies also do not suffer from this artefact because the emission lines are directly measured towards H\textsc{ii} regions.

To estimate the errors in the metallicity gradients, we adopt a bootstrapping approach similar to \citet{Kewley:2010kx}, \citet{Rupke:2010fk} and \citet{Rich:2012oq}. We randomly draw from the measurements the same number of data points but with replacement, and perform an unweighted least-squares linear fit with the drawn data. In each galaxy, this process is repeated 1,000 times and each fit result is recorded. The median and standard deviation of the slopes and intercepts are considered as the best estimates of the metallicity gradient.

\section{Result}\label{sec-result}

\subsection{Metallicity gradients of individual galaxies}

In Figure~\ref{califa_gradient} and Figure~\ref{s12_gradient}, we present 4 metallicity gradients measured in two CALIFA and two S12 galaxies, respectively. The two different metallicites are shown in the first two panels, and the last panels show radial profiles of the ionisation parameter. In the first two panels, straight lines indicate the best fits of the gradients, and dashed lines indicate 1$\sigma$ errors as propagated using analytic expressions with bootstrapped errors. The rest of the CALIFA and S12 galaxies are presented in Figures~\ref{more_califa} and \ref{more_s12} in Appendix~\ref{appendix}. 

The metallicity gradients and the radial profiles of the ionisation parameter of our 10 WiFeS galaxies are also presented in Figure~\ref{wifes_gradient}. The metallicity of the majority of WiFeS galaxies have metallicities $\rm 12+\log(O/H) < 8.4$ (in KD02 scale), where the conversion from O3N2 to N2O2 is difficult to determine, and therefore we are only able to derive their N2O2 metallicity gradients. These galaxies are excluded from the rest of the comparisons between two metallicity calibrations, but we include them later while comparing metallicity gradients of different galaxies (Section~\ref{sec-metallicity-gradients-in-field-star-forming-galaxies}).

\begin{figure}
\centering
\includegraphics[width=8.5cm]{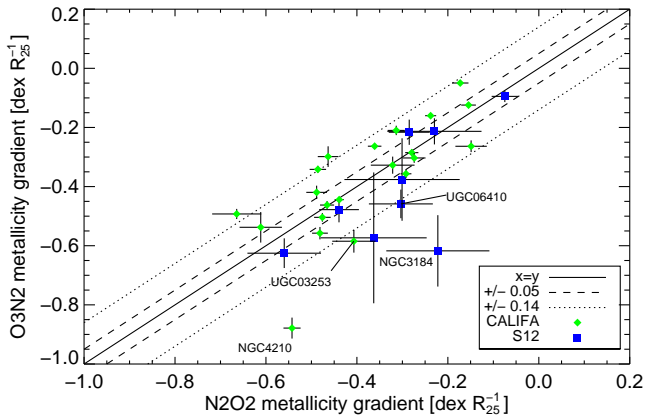} 
\caption{{\it Left:} Comparison between the metallicity gradients (${\rm dex}~R_{25}^{-1}$) derived using the O3N2 and the N2O2 diagnostics (see Section~\ref{sec-metallicity}). The metallicity gradients of the majority of the galaxies  (33\%/73\%) agree within $\pm0.05/\pm0.14~{\rm dex}~R_{25}^{-1}$. The metallicity gradients of the four outliers labeled are shown in Figure~\ref{califa_gradient} and Figure~\ref{s12_gradient}. }\label{compare_gradient}
\end{figure}

In Figure~\ref{compare_gradient}, we compare metallicity gradients derived from O3N2 and from N2O2 for the CALIFA and S12 galaxies. All the galaxies exhibit negative metallicity gradients in both calibrations, and 33\%/73\% of the galaxies  agrees within $\pm0.05/\pm0.14~{\rm dex}~R_{25}^{-1}$. These values can be considered as the level of residual systematics in the metallicity calibrations. Noticeably there are several outliers well below the one-to-one line which we label in Figure~\ref{compare_gradient}. These objects could provide insight into the cause of the disagreement between the two calibrations. We further investigate the cause of this discrepancy in the following two subsections. 

\floatplacement{figure}{!b}

\subsection{The effect of ionisation parameter}\label{sec-the-effect-of-ionisation-parameter}
\floatplacement{figure}{H}
\begin{figure*}
\centering
\includegraphics[width=17cm]{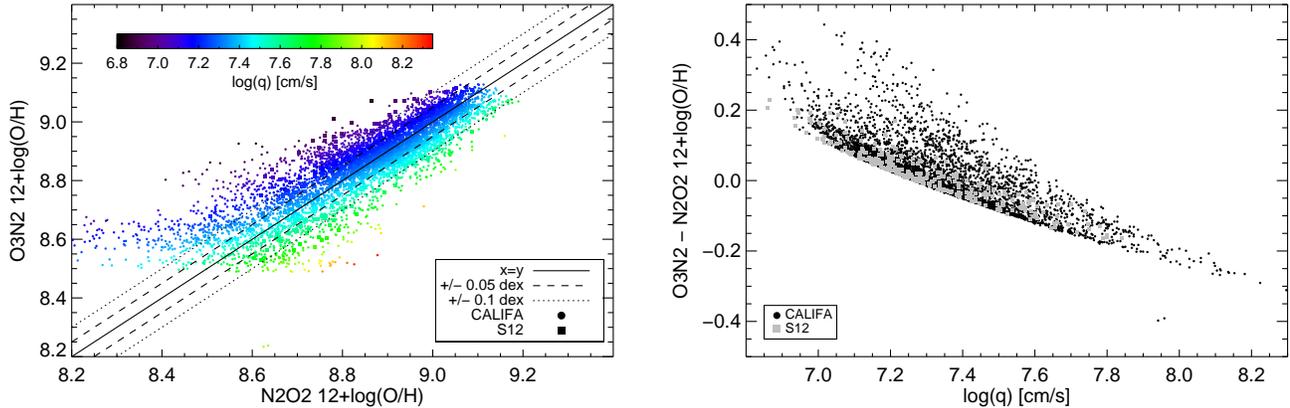} 
\caption{{\it Left:} Comparison between the metallicities derived using the O3N2 and N2O2 diagnostics (see Section~\ref{sec-metallicity}). Each CALIFA data point is an IFU spaxel, and each S12 data point is an H\textsc{ii} region extracted from IFU data. Data points are colour-coded with their corresponding ionisation parameters (see Section~\ref{sec-ionisation-parameter}). A total of 57\%/83\% of the data points agrees within $\pm0.05/0.1~\rm dex$. The degree of disagreement correlates strongly with the ionisation parameter. {\it Right:} Difference between the metallicities derived using the O3N2 and N2O2 diagnostics versus the ionisation parameter. A strong anti-correlation between the two quantities is obvious. More discussion about this discrepancy of metallicities is provided in Section~\ref{sec-the-effect-of-ionisation-parameter}, and the impact on measuring metallicity gradients in Section~\ref{sec-discrepancies-among-metallicity-gradients}.  }\label{compare_z}
\end{figure*}

In the left panel of Figure~\ref{compare_z}, we compare the metallicities derived using the O3N2 and N2O2 calibrations. Each point represents one measurement, i.e., one spaxel from the CALIFA galaxies or one H\textsc{ii} region from the S12 galaxies. We colour-code the points by log($q$) derived using the [\ion{O}{iii}]/[\ion{O}{ii}] (KK04) calibration. The left panel of Figure~\ref{compare_z} clearly demonstrates that superficially the two metallicities agree reasonably well; 57\%/83\% of the metallicities agree within $\pm0.05/0.1~\rm dex$. Similar comparisons were also carried out by \citet{Rupke:2010fk} where comparable scattering also exists.

In the left panel of Figure~\ref{compare_z}, the scattering around the one-to-one line is not random, but correlates with ionisation parameter. Spaxels or slits with high ionisation parameter have higher N2O2 metallicities than O3N2 metallicities. Those with low ionisation parameter have lower N2O2 metallicity than O3N2 metallicities. In the right panel of Figure~\ref{compare_z}, we show the difference between metallicities derived with the O3N2 and N2O2 ratios versus ionisation parameter. The differences between the two diagnostics can be up to 0.2 -- 0.4~dex at extreme values of ionisation parameter ($\log(q) < 7.0\rm~cm~s^{-1}$  or $\log(q) > 8.2\rm~cm~s^{-1}$ ). At $\log(q) \gtrsim (\lesssim) 7.3\rm~cm~s^{-1}$, the O3N2 diagnostic gives higher (lower) metallicity than the N2O2 diagnostic. Clearly, ionisation parameter is the cause of the discrepancy. As we emphasised earlier, O3N2 is calibrated empirically without taken into account the change of ionisation parameter. A theoretical calibration of O3N2 will be presented in Kewley et al. (in preparation) and will reconcile this discrepancy with new stellar population synthesis and photoionisation models.

\subsection{Discrepancies among metallicity gradients }\label{sec-discrepancies-among-metallicity-gradients}

We now return to discuss the cause of the discrepancies in the metallicity gradients shown in Figure~\ref{compare_gradient}. 

Figure~\ref{califa_gradient} (and Figure~\ref{s12_gradient}) show metallicity gradients for two CALIFA (two S12 galaxies) that are labeled as outliers in Figure~\ref{compare_gradient}. These galaxies have steeper O3N2 than N2O2 metallicity gradients because, at large radii, O3N2 metallicities are systematically lower than N2O2 metallicities. As shown in Figure~\ref{compare_z}, lower O3N2 than N2O2 metallicities naturally arises when ${\log}(q) \gtrsim 7.3\rm~cm~s^{-1}$. Indeed in the third panels of Figure~\ref{califa_gradient} and Figure~\ref{s12_gradient}, these galaxies typically have ${\log}(q) \gtrsim 7.3\rm~cm~s^{-1}$ at large radii. These galaxies all show indications of smooth rising of ionisation parameter from their centres to outskirts, implying a continuous radial change of their properties of the ionising radiation. The higher ionisation parameters could be caused by the more active star formation activities with different distributions of molecular gas \citep{Dopita:2014qy}.

In extreme cases, the differences in metallicity gradient measured with  N2O2 and O3N2 can be up to a $\sim0.4~{\rm dex}~R_{25}^{-1}$ (e.g., NGC3184 in Figure~\ref{s12_gradient}). Similar findings are also reported in \citet{Rupke:2010fk}. These results have important implications for metallicity gradient studies at high redshift, where typically only  [\ion{N}{ii}]~$\lambda\lambda$6548,6583 and H$\alpha$ are available \citep[{in some cases also  [\ion{O}{iii}]~$\lambda\lambda$4959,5007 and H$\beta$, e.g.,}][]{Cresci:2010fk,Yuan:2011qy,Jones:2010uq,Jones:2013kx,Queyrel:2012qe,Swinbank:2012ve}. While all the diagnostics using these four lines, i.e. [\ion{N}{ii}]~$\lambda$6583/H$\alpha$ and O3N2, are sensitive to the change of ionisation parameter, quantitative interpretation of metallicity gradients should bear in mind the potential impact of ionisation parameter gradients in galaxies.

For galaxies not labeled in Figure~\ref{compare_gradient}, we do not find obvious signs of a correlation between ionisation parameter and radius. The inconsistency in metallicities derived using the O3N2 and N2O2 diagnostics does not correlate with radius. When measuring metallicity gradients in these galaxies, the net effect is merely to increase the uncertainties in the measurements, rather than biasing the measurements in any systematic way. This is essentially the cause of the small scatter ($\lesssim0.1~{\rm dex}~R_{25}^{-1}$) in Figure~\ref{compare_gradient}.

\subsection{Metallicity gradients in field star-forming galaxies}\label{sec-metallicity-gradients-in-field-star-forming-galaxies}
After investigating the systematics induced by the variation of ionisation parameter, we adopt the N2O2 measurements as our final metallicity gradients and we now compare metallicity gradients in field star-forming galaxies with their stellar mass and {\it B}-band luminosity.

\floatplacement{figure}{!t}

\begin{figure}
\centering
\includegraphics[width=8.5cm]{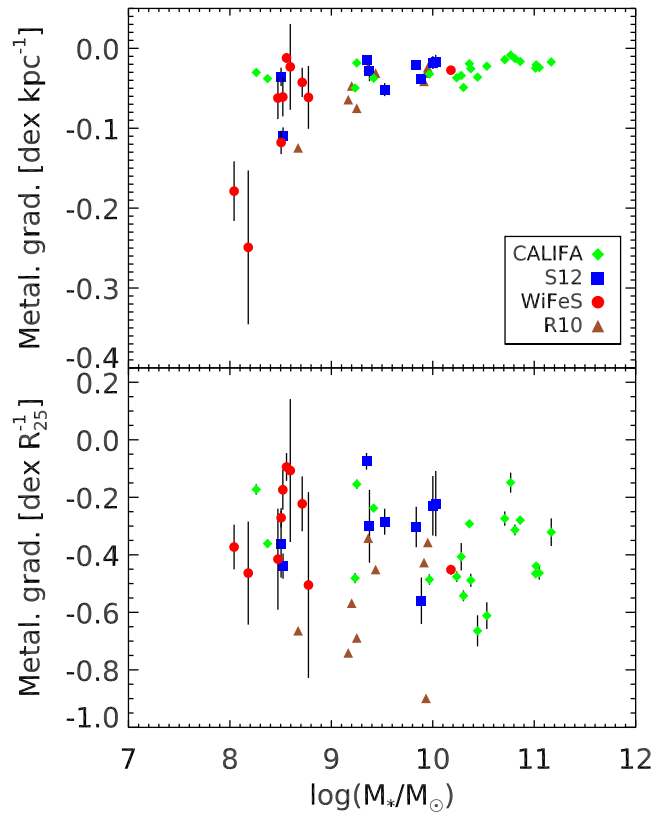} 
\caption{Metallicity gradient versus stellar mass when the metallicity gradients are measured in $\rm dex~kpc^{-1}$ (upper panel) and in ${\rm dex}~R_{25}^{-1}$ (lower panel). More details are discussed in Section~\ref{sec-metallicity-gradient-stellar-mass}. }\label{gradient-mass}
\end{figure}

\subsubsection{Metallicity gradient - stellar mass}\label{sec-metallicity-gradient-stellar-mass}

The upper and lower panels of Figure~\ref{gradient-mass} show the metallicity gradient versus stellar mass of our four samples in units of $\rm dex~kpc^{-1}$ and dex~$R_{25}^{-1}$, respectively. In the upper panel, the metallicity gradients appear to depend on stellar mass where (1) low mass galaxies have on average a steeper metallicity gradient, and (2) are more diverse in the steepness of metallicity gradients compared to high-mass galaxies. To quantify the dependency, we split the sample into two mass bins with roughly equal numbers of galaxies, i.e. a high-mass bin (${\log}(M_*/M_\odot)>9.6$; $N_{gal} = 24$)  and a low-mass bin (${\log}(M_*/M_\odot)<9.6$; $N_{gal}=25$). We compute the bootstrapped means and standard deviations of the metallicity gradients. The results are tabulated in Table~\ref{table-benchmark}. We indeed find that the low-mass galaxies have a steeper mean metallicity gradient at 3.4$\sigma$ level than the high-mass galaxies, and also a larger standard deviation of the metallicity gradients at 3.7$\sigma$ level. 

\floatplacement{figure}{!t}
\begin{figure}
\centering
\includegraphics[width=8.5cm]{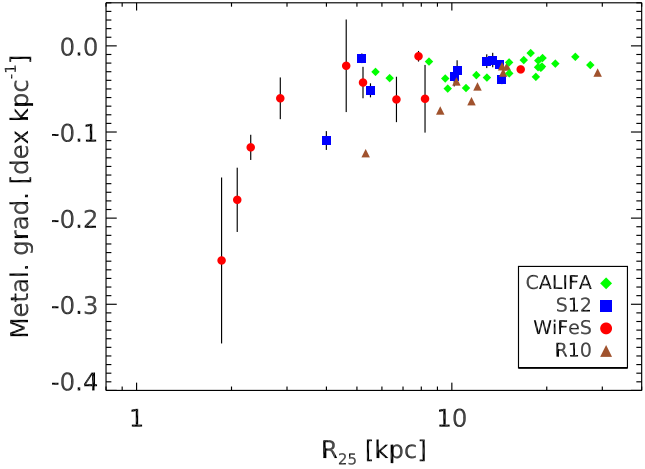} 
\caption{Metallicity gradients ($\rm dex~kpc^{-1}$) versus $R_{25}$ in kpc. A correlation between the two quantities can be seen. More details are discussed in Section~\ref{sec-metallicity-gradient-stellar-mass}.  }\label{gradient-kpc-r25}
\end{figure}

\floatplacement{figure}{!t}
\begin{figure}
\centering
\includegraphics[width=8.5cm]{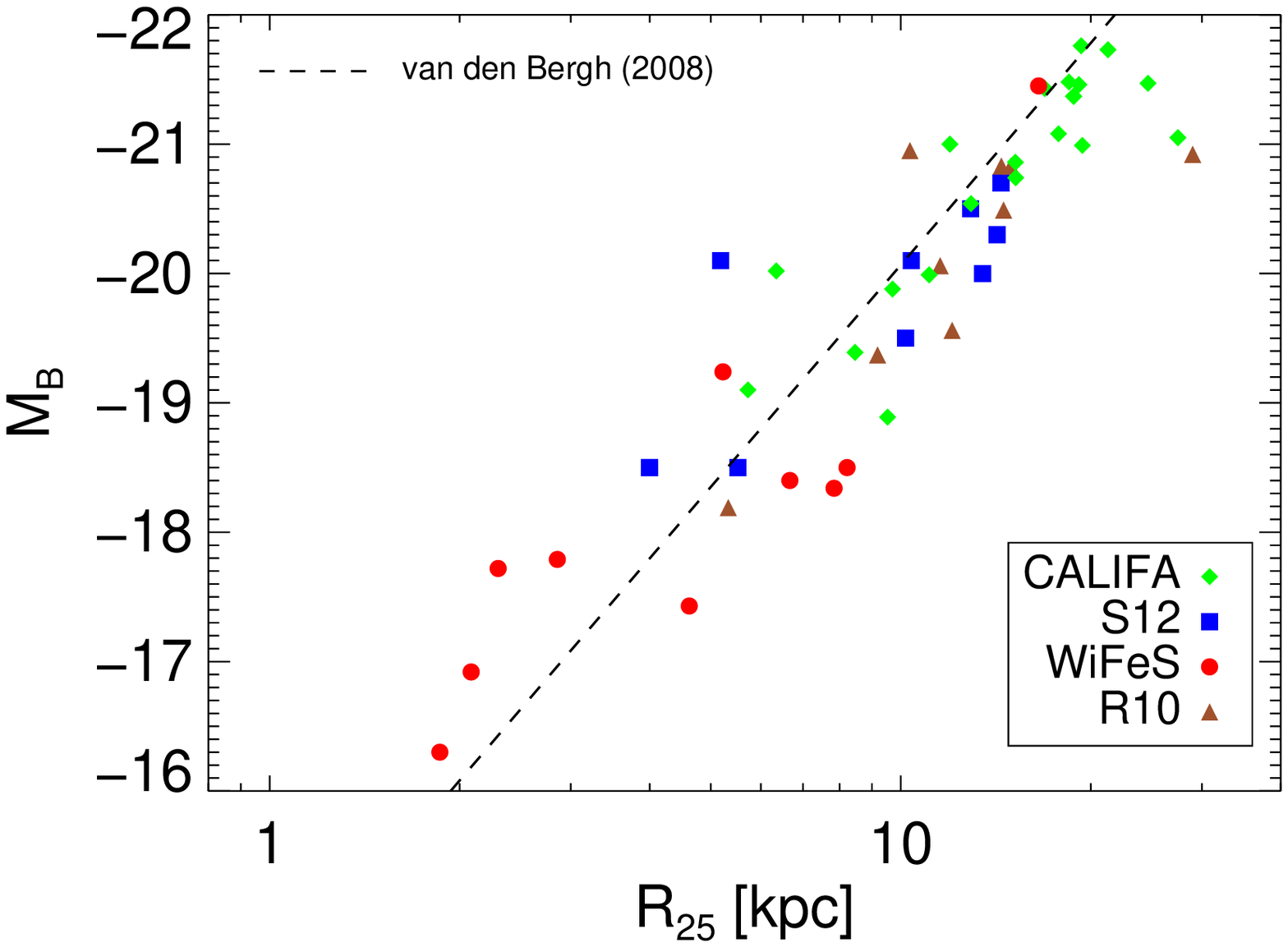} 
\caption{ {\it B}-band luminosity versus $R_{25}$ in kpc. The dashed line indicates the luminosity-size relation by \citet{van-den-Bergh:2008rt}. \label{r25_mb}}
\end{figure}

When the metallicity gradients are normalised to the galaxy sizes (i.e. dex~$R_{25}^{-1}$), the lower panel of Figure~\ref{gradient-mass} does not support any clear dependency of metallicity gradient on stellar mass. The difference between the means is only at 1.2$\sigma$ level, with low-mass galaxies having slightly flatter metallicity gradients than high-mass galaxies. The standard deviations of the metallicity gradients are virtually identical (the difference is only 0.6$\sigma$).

The different dependency of metallicity gradient on stellar mass while measuring the metallicity gradient in absolute scale (kpc) or relative scale ($R_{25}$) can be understood as a size effect. If galaxies with steeper $\rm dex~kpc^{-1}$ metallicity gradients are smaller in their physical sizes (small $R_{25}$), then the steep $\rm dex~kpc^{-1}$ metallicity gradients would be compensated when the galaxy sizes are taken into account. Figure~\ref{gradient-kpc-r25} shows the $\rm dex~kpc^{-1}$ metallicity gradient versus galaxy size $R_{25}$ of our samples. Indeed, galaxies with steeper metallicity gradients generally have smaller $R_{25}$ than galaxies with shallower metallicity gradients. The Spearman rank correlation coefficient between $R_{25}$ and the $\rm dex~kpc^{-1}$ metallicity gradients is 0.6, which is different from zero (i.e. no correlation) at a significance of $7.5\times10^{-6}$. In Figure~\ref{r25_mb}, we demonstrate that our samples fall on the luminosity-size relation \citep{van-den-Bergh:2008rt}, i.e. low luminosity galaxies have smaller $R_{25}$ than high luminosity galaxies, indicating that the steep $\rm dex~kpc^{-1}$ metallicity gradients in low-mass galaxies could be associated with their small physical sizes. These results imply that the evolution of metallicity gradients is closely related to the growth of galaxy size. 

\subsubsection{Metallicity gradient - $M_{B}$}\label{sec-metallicity-gradient-mb}

\begin{figure}
\centering
\includegraphics[width=8.5cm]{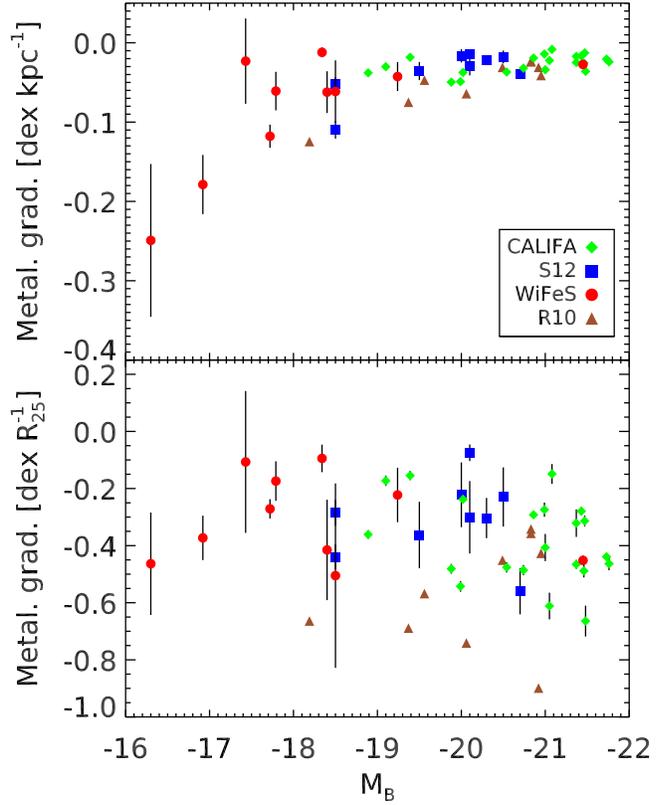} 
\caption{Metallicity gradient versus absolute {\it B}-band magnitude when the metallicity gradients are measured in $\rm dex~kpc^{-1}$ (upper panel) and in ${\rm dex}~R_{25}^{-1}$ (lower panel). More details are discussed in Section~\ref{sec-metallicity-gradient-mb}.  }\label{gradient-mb}
\end{figure}

We now compare metallicity gradients with the absolute {\it B}-band magnitudes, $M_{B}$, for our four samples. The absolute {\it B}-band magnitude is used as a proxy for mass in metallicity studies where multi-band photometry is not available. \citet{Rubin:1984fk} first showed that metallicity is correlated with luminosity in disk galaxies. Further investigations solidified the luminosity-metallicity correlation in nearby disk galaxies \citep{Bothun:1984fk,Wyse:1985fk,Skillman:1989lr,Vila-Costas:1992fj,Zaritsky:1994lr,Garnett:2002fj}. We emphasise that optical luminosity is not always a reliable surrogate for the stellar mass of a galaxy because optical luminosities are sensitive to the current SFR and are affected by dust.  Near-infrared luminosities can be influenced by the age of the stellar population of a galaxy.  Since this quantity is widely adopted in earlier studies, we present our measurements below for comparison.

In Figure~\ref{gradient-mb}, we present metallicity gradient versus $M_{B}$ of our four samples. Metallicity gradients are shown in both $\rm dex~kpc^{-1}$ (upper panel) and ${\rm dex}~R_{25}^{-1}$ (lower panel). Similarly, we split the sample into a high luminosity bin ($M_B<-20.1$; $N_{gal} = 24$) and a low luminosity bin ($M_B>-20.1$; $N_{gal} = 25$), and compute the bootstrapped means and standard deviations of the metallicity gradients (Table~\ref{table-benchmark}). We reach the similar conclusions, as in the comparisons with stellar mass, that (1) low luminosity galaxies have a steeper mean $\rm dex~kpc^{-1}$ metallicity gradient ($3.4\sigma$), and (2) low luminosity galaxies have a larger standard deviation of $\rm dex~kpc^{-1}$ metallicity gradients ($3.5\sigma$). When the galaxy sizes are taken into account, i.e. dex~$R_{25}^{-1}$, the low and high luminosity galaxies have very similar mean metallicity gradients ($1.4\sigma$) and standard deviations ($1.8\sigma$). These findings are in agreement with previous studies \citep{Vila-Costas:1992fj,Garnett:1997fj,Prantzos:2000vn}. The lack of correlations between the metallicity gradient (in dex per scale-length) and macroscopic properties such as stellar mass and {\it B}-band magnitude might imply that the relationships between these parameters are more complex than simple correlations, a hypothesis first suggested by \citet{Zaritsky:1994lr} and recently investigated by \citet{Pilyugin:2014lr}.

\citet{Few:2012gf} performed cosmological zoom-in simulations that yield metallicity gradients, disk scale-lengths, {\it B}-band magnitudes, and stellar masses of 19 galaxies in field and loose group environments. The simulations focused on Milky Way-mass galaxies spanning stellar mass and {\it B}-band magnitude ranges of $10.4<{\log}(M_*/M_\odot)<11.1$ and $-19.7>M_B>-21.7$, respectively, which correspond to the high-mass and high-luminosity ends of our samples. \citet{Few:2012gf} found no significant differences in metallicity gradients between the galaxies in field and loose group. The overall metallicity gradients are $-0.046\pm0.013\rm~dex~kpc^{-1}$ and $-0.40\pm0.13{\rm~dex}~R_{25}^{-1}$ (mean $\pm$ standard deviation). Here, we convert the disk scale lengths $R_d$ reported by \citet{Few:2012gf} to $R_{25}$ assuming that the exponential disks have the canonical central surface brightness for normal spirals of 21.65 magnitude/arcsec$^2$ \citep{Freeman:1970fk}. The overall metallicity gradients by  \citet{Few:2012gf} are consistent with our results in the high-mass and high-luminosity ends. Similar simulations in the future targeting lower masses and luminosities could provide valuable constraints for the different prescriptions built into the simulations.


\begin{table}
 \caption{A local benchmark gradient}
 \label{table-benchmark}
 \begin{tabular}{ccc}
  \hline
  & Mean & Standard deviation\\
\hline
\multicolumn{3}{c}{$\rm dex~kpc^{-1}$}\\
\hline
${\log}(M_*/M_\odot)>9.6$ & $-0.026\pm0.002$  & $0.010\pm0.001$  \\
${\log}(M_*/M_\odot)<9.6$ & $-0.064\pm0.011$  & $0.054\pm0.012$ \\
$M_B < -20.1$ & $-0.025\pm0.002$  & $0.008\pm0.001$  \\
$M_B > -20.1$ & $-0.063\pm0.011$  & $0.053\pm0.013$  \\
\hline
\multicolumn{3}{c}{${\rm dex}~R_{25}^{-1}$} \\
\hline
${\log}(M_*/M_\odot)>9.6$ & $-0.42\pm0.03$ & $0.16\pm0.03$ \\
${\log}(M_*/M_\odot)<9.6$ & $-0.36\pm0.04$ & $0.18\pm0.02$ \\
$M_B < -20.1$ & $-0.40\pm0.03$  & $0.12\pm0.02$  \\
$M_B > -20.1$ & $-0.34\pm0.03$  & $0.17\pm0.02$  \\
 All\tablenotemark{a} & $-0.39\tablenotemark{a}$ & $0.18\tablenotemark{a}$ \\
\hline
\end{tabular}
\medskip 

Means, standard deviations and the associated errors are derived from bootstrapping.\\
$\rm ^a$Mean and standard deviation of all the galaxies (i.e. not from bootstrapping). See Figure~\ref{gradient-benchmark}. 
\end{table}

\section{A local benchmark gradient}\label{sec-benchmark-gradient}

\begin{figure}
\centering
\includegraphics[width=8.3cm]{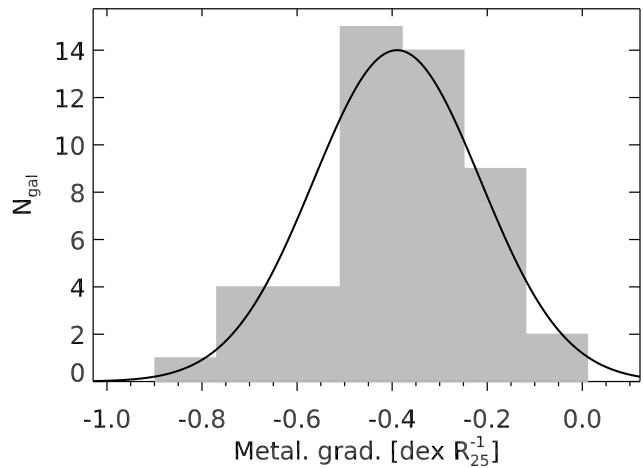} 
\caption{Distribution of the 49 metallicity gradients. The overall mean and standard deviation of the metallicity gradients are $-0.39\pm0.18~{\rm dex}~R_{25}^{-1}$ (i.e. the benchmark metallicity gradient). The black curve indicates a Gaussian with these characteristic values, i.e. not a fit to the distribution. A one-sided Kolmogorov-Smirnov test yields a probability of 78\% for the observed distribution to be drawn from the back curve. }\label{gradient-benchmark}
\end{figure}

We provide a local benchmark of metallicity gradients inspired by the uniformity of the ${\rm dex}~R_{25}^{-1}$ metallicity gradients. We believe the local benchmark gradient will be useful for comparison with metallicity gradients measured at high redshift. In Figure~\ref{gradient-benchmark}, we present the distribution of the 49 measured metallicity gradients; and we summarise the mean and standard deviation in Table~\ref{table-benchmark}. The benchmark metallicity gradient measures $-0.39\pm0.18~{\rm dex}~R_{25}^{-1}$ (mean $\pm$ standard deviation). A one-sided Kolmogorov-Smirnov test yields a 78\% probability for the observed distribution to be drawn from the normal distribution shown as the black curve in Figure~\ref{gradient-benchmark}.  We note that the difference between the O3N2 and N2O2 metallicity gradients is typically $0.14~{\rm dex}~{R_{25}^{-1}}$ (Figure~\ref{compare_gradient}),  comparable to the standard deviation of $0.18~{\rm dex}~{R_{25}^{-1}}$ in the benchmark gradient. This remarkably small difference suggests that the intrinsic spread of the metallicity gradients could be even tighter than $0.18~{\rm dex}~{R_{25}^{-1}}$ although precisely quantifying the tightness is non-trivial due to the various systematic effects. 

We note that in \mbox{Figures~\ref{gradient-mass} and \ref{gradient-mb}}, the R10 sample appears to have lower metallicity gradients than the other three samples. A one-sided Kolmogorov-Smirnov test comparing the R10 sample to the benchmark gradient, and a two-sided Kolmogorov-Smirnov test comparing the R10 sample to the whole sample both reveal moderately low but non-zero probability ($10\%$) for the two distributions to be the same. It is possible that the discrepancy is due to low number statistics. Furthermore, a two-sided Kolmogorov-Smirnov test comparing the full sample to that with the R10 galaxies excluded yields a high probability (99\%) for the two distributions to be drawn from the same parent distribution. We conclude that the discrepancy is only apparent and does not change our results. 

S12 measured metallicity gradients in 25 face-on spirals and found a common metallicity gradient of $-0.12\pm0.11~{\rm dex}~R_e^{-1}$ ($\rm median \pm standard\ deviation$). \citet{Sanchez:2014fk} expanded the study to 193 galaxies with the CALIFA survey, and found a very similar common metallicity gradient of $-0.10\pm0.09~{\rm dex}~R_e^{-1}$. \citet{Sanchez:2014fk} showed that, with their large sample size, the metallicity gradients are independent of morphology, incidence of bars, absolute magnitude and mass, a result that had also been hinted by S12. \citet{Sanchez:2014fk} found that the only clear correlation is between merger stage and metallicity gradient, where the slope is flattened as merger progresses \citep[see also][]{Kewley:2010kx,Rupke:2008fk,Rupke:2010lr,Rupke:2010fk,Torrey:2012ai,Rich:2012oq}. 

To compare the common metallicity gradient by \citet{Sanchez:2014fk} with our benchmark gradient, one must take into account the different scale-lengths, i.e. $R_e$ and $R_{25}$, and metallicity calibrations (\citealt{Sanchez:2014fk} and S12 both adopted the O3N2/PP04 calibration). Assuming again an exponential disk with the canonical central surface brightness for normal spirals \citep[i.e. 21.65 magnitude/arcsec$^2$;][]{Freeman:1970fk} and the empirical conversion between the O3N2/PP04 and N2O2/KD02 metallicities \citep{Kewley:2008qy}, we can convert the common metallicity gradient by \citet{Sanchez:2014fk} to $-0.20\pm0.18~{\rm dex}~R_{25}^{-1}$, which is about a factor of two shallower than our benchmark gradient. The difference could be caused by the presence of close pairs and mergers in their sample. In addition, systematic errors such as the metallicity calibrations, variations of the ionisation parameter, and flattening of metallicity gradient at larger radii could all affect the measured metallicity gradients.

\begin{figure}
\centering
\includegraphics[width=8.6cm]{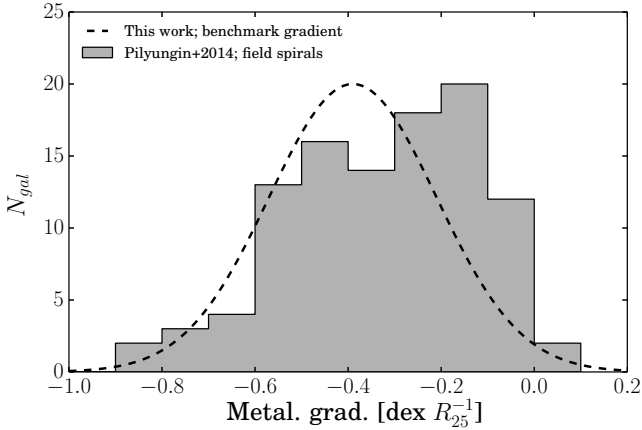}
\caption{Comparison between our benchmark gradient (dashed curve) and the metallicity gradients from 104 field spiral galaxies published by \citet{Pilyugin:2014uq}. Our benchmark gradient is $-0.39\pm0.18~{\rm dex}~R_{25}^{-1}$ (mean $\pm$ standard deviation; Figure~\ref{gradient-benchmark}). The metallicity gradients of the field spiral galaxies from \citet{Pilyugin:2014uq} measure $-0.32\pm0.20~{\rm dex}~R_{25}^{-1}$. }\label{p2014_dist}
\end{figure}

\citet{Pilyugin:2014uq} complied more than 3,000 published spectra of H\textsc{ii} regions and adopted the calibration proposed by \citet{Pilyugin:2012fk} to measure the metallicity gradients of 130 nearby late-type galaxies. Using 104 of their field spiral galaxies (i.e. excluding mergers and close pairs), we constrain a mean and standard deviation of the metallicity gradients of $-0.32\pm0.20~{\rm dex}~R_{25}^{-1}$. These values are consistent with our benchmark gradient (see Figure~\ref{p2014_dist} for a comparison), but the distribution from their sample do not match our benchmark gradient, implying residual systematic errors.

\section{Why is there a benchmark gradient?}\label{sec-why-benchmark-gradient}

The existence of a common metallicity gradient when expressed with respect to some scale-lengths had long been suggested \citep[e.g.,][]{Zaritsky:1994lr,Garnett:1997fj,Vila-Costas:1992fj}, and is further solidified by recent observations with integral field spectroscopy on large samples (S12; \citealt{Sanchez:2014fk}). The common slope implies that all disk galaxies went through very similar chemical evolution when building up their disks, presumably in an inside-out fashion \citep{Sanchez:2014fk}. We show that our benchmark gradient is closely related to the growth of galaxy size, indicating that the chemical richness of disk galaxies co-evolves with the increase in their spatial dimensions \citep[ses also][]{Prantzos:2000vn}. 

Since the measured metallicity is the ratio of oxygen to hydrogen atoms, to the zeroth-order the metallicity ought to be a strong function of the stellar-to-gas mass ratio. The stellar mass traces the total amount of metals produced through the formation of stars that drive nucleosynthesis; and the gas mass serves as the normalisation for the definition of metallicity. In the simplest case, known as the ``closed-box'' model, the chemical evolution began with pristine gas and experienced no subsequent mass exchange with the material outside the box. This classical closed-box model \citep{Searle:1972ul,Pagel:1975zr} already encapsulates the close link between the metallicity, $Z$, and the observed stellar mass to gas mass ratio, $M_{*o}/M_{g}$:
\begin{eqnarray}\label{main-eq-closed-box}
Z(t) & = & {y \over 1-R } \ln \left[ 1 + {M_{*o}(t)\over M_{g}(t)}  \right] \\
       & = & {y \over 1-R } \ln \left[ {1\over \mu_g (t)}  \right].
\end{eqnarray}
Here, $Z$ is the mass ratio instead of the number ratio adopted in $\rm 12 + \log(O/H)$, $y$ is the nucleosynthesis yield and $R$ is the stellar returned mass fraction. The ``observed'' stellar mass, $M_{*o}$, takes into account the mass loss through stellar winds described by the returned mass fraction, i.e. $M_{*o} = (1-R) M_*$, where the time derivative of $M_*$ is the star formation rate. 
The gas fraction $\mu_g$ is defined as
\begin{equation}
\mu_g(t) = { M_{g}(t) \over  M_{g}(t) +  M_{*o}(t)}. 
\end{equation}
Such simple picture, however, is usually complicated by gas inflows (e.g., accretion and merger) and outflows (e.g., AGN and starburst feedback) that can remove or replenish both gas and metals to the ISM and break down the closed-box assumption. 

It is possible to put constraints on the inflow and outflow history by studying the global metallicity of star-forming galaxies using the mass-metallicity relation (e.g., \citealt{Spitoni:2010gf,Peeples:2011kx,Zahid:2012ys,Lilly:2013qf,Peeples:2014fj}; see also \citealt{Dave:2011qy}). Global studies of metallicity suggest a close relation between metallicity, gas mass, and stellar mass, and the potential of the co-evolution of these three quantities following a simple, universal relation between metallicity and stellar-to-gas mass ratio \citep[][]{Zahid:2014uq,Ascasibar:2014fk}. Similar universal relations may also exist on spatially resolved scales (\citealt{Ascasibar:2014fk}; see also \citealt{Rosales-Ortega:2012lq}).

\begin{figure}
\centering
\includegraphics[width=8.6cm]{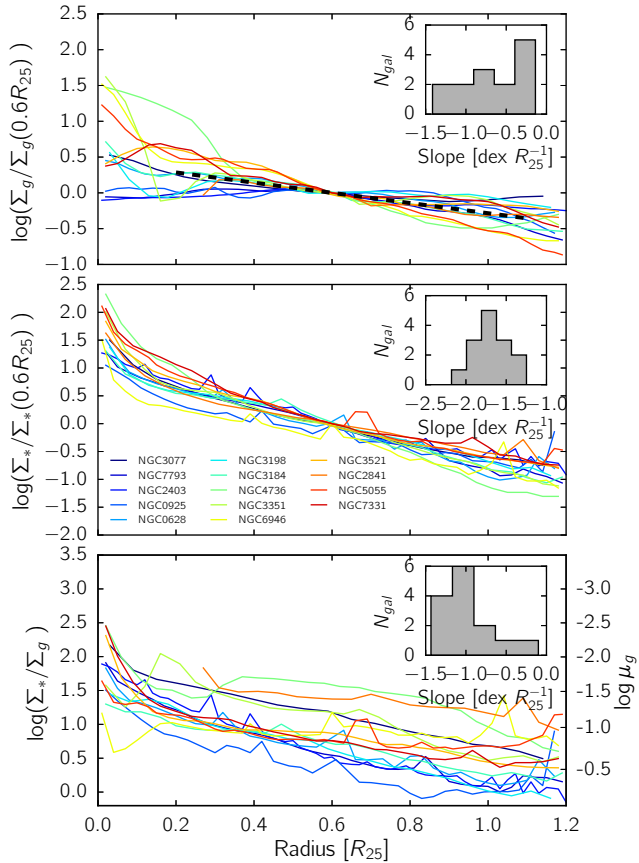}
\caption{Gas (atomic + molecular; top panel), stellar (middle panel), and stellar-to-gas (bottom panel) surface density profiles of 14 field spiral galaxies from \citet{Leroy:2008qy}. The radial distance is expressed in terms of $R_{25}$. The gas and stellar profiles are normalised at 0.6$R_{25}$, and the stellar-to-gas profiles are not normalised. The dashed line in the top panel indicates the best-fit universal gas profile from \citet{Bigiel:2012qy}. 
The insets show the distributions of the slopes measured by fitting straight lines to the logarithmic profiles using data at $r>0.2R_{25}$. }\label{ms_mg_profile}
\end{figure}

If such relation exists, the common metallicity gradient implies a close link between the radial profiles of the stars and the gas. Indeed, carbon monoxide and H\textsc{i} 21~cm observations in nearby spiral galaxies reveal that the neutral (molecular and atomic) gas surface density profiles, $\Sigma_{\rm g}(r)$, exhibit a tight universal profile \citep{Bigiel:2012qy}. When the gas surface density profiles of the individual galaxies are expressed in terms of $R_{25}$ and normalised to a transition radius where the molecular and atomic gas have the same surface densities, \cite{Bigiel:2012qy} show that the overall surface density of all the galaxies (at $r>0.2R_{25}$) follows a simple exponential profile. The exponential profile has a logarithmic slope of $-0.71~{\rm dex}~{R_{25}^{-1}}$ and a very small bootstrapped error of the mean of $0.06~{\rm dex}~{R_{25}^{-1}}$ (see their figure~3). In Figure~\ref{ms_mg_profile}, we show the gas, the stellar ($\Sigma_*(r)$), and the stellar-to-gas surface density profiles of 14 field spiral galaxies from \citet[][$9.3<{\log}(M_*/M_\odot)<10.9$; $R_{25} = 3 \mbox{--} 20~{\rm kpc}$; c.f. Figures~\ref{gradient-mass} and \ref{gradient-kpc-r25}]{Leroy:2008qy}. All the profiles are expressed in terms of $R_{25}$, and we normalise the gas and stellar surface density profiles at 0.6$R_{25}$. Clearly, the gas, stellar, and stellar-to-gas surface density all follow simple radial profiles with common slopes. We fit simple exponential profiles to each galaxy using data at $r>0.2R_{25}$ to constrain the logarithmic slopes, and we find that the distributions of the slopes have means $\pm$ standard deviations of $-0.69\pm0.40~{\rm dex}~{R_{25}^{-1}}$ for the gas, $-1.68\pm0.23~{\rm dex}~{R_{25}^{-1}}$ for the stars, and $-0.98\pm0.35~{\rm dex}~{R_{25}^{-1}}$ for the stellar-to-gas surface density. 

\begin{figure}
\centering
\includegraphics[width=8.6cm]{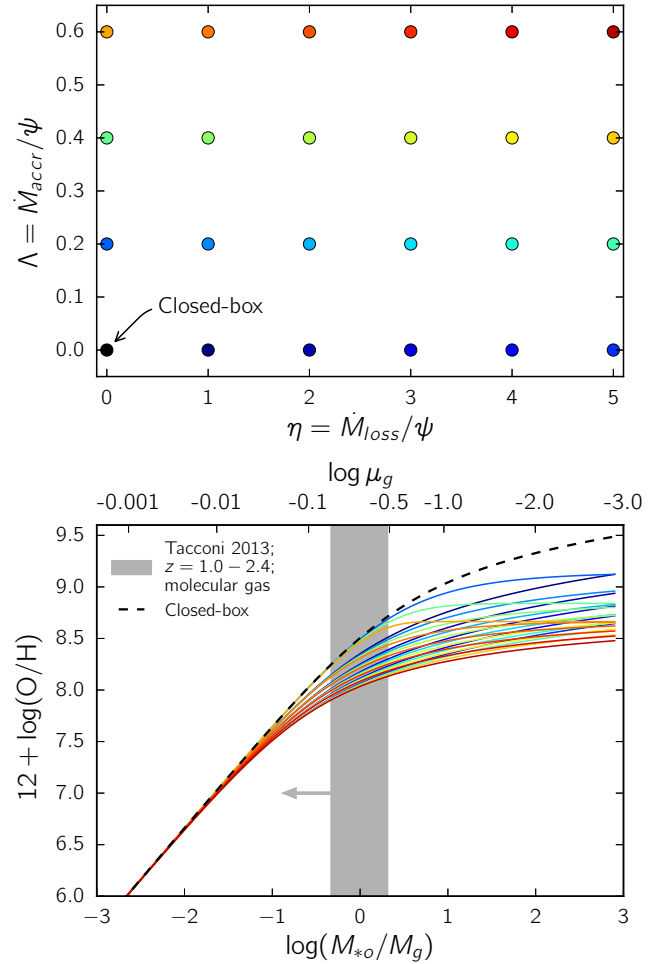}
\caption{Examples of our chemical evolution models. The upper panel shows the $\eta\mbox{-}\Lambda$ plane. Each colour point corresponds to one model determined by the set of mass loading and mass accretion factors. The lower panel show the corresponding relationship between the metallicity and stellar-to-gas mass ratio. The black dot at $(\eta,\Lambda)=(0,0)$ in the upper panel is the closed-box model; the corresponding curve in the lower panel is shown as the dashed curve. The gray band marks the $\pm1\sigma$ range of the 73 massive star-forming galaxies measured by \citet{Tacconi:2013ly}. As only the molecular gas was measured, not the atomic gas, the range represents an upper limit. We note that $M_{*o}$ denotes the ``observed'' stellar mass taken into account the stellar mass return. }\label{modelExamples}
\end{figure}

These tight universal profiles, in particular the stellar-to-gas mass ratio, may govern the common metallicity gradient in disk galaxies. We use simple chemical evolution models to quantitatively address the relation between metallicity, and stellar-to-gas mass ratio. The models we adopted are special cases of the more general derivations by \cite{Recchi:2008ly}. Similar models have been applied to global metallicities of galaxies \citep[e.g.,][]{Spitoni:2010gf,Dayal:2013ul,Lilly:2013qf,Pipino:2014ve}. The analytical models consider, for a given volume element, metal production by stars, stellar mass return, inflows, and outflows, under the assumption that the gas is well-mixed and the mass return from stars is recycled instantaneously. 
The inflows and outflows are described through two critical parameters: the mass loading factor 
\begin{equation}
\eta \equiv {\dot{M}_{loss} \over \psi}
\end{equation}
and the mass accretion factor 
\begin{equation}
\Lambda \equiv {\dot{M}_{accr} \over \psi},
\end{equation}
where $\dot{M}_{loss}$ and $\dot{M}_{accr}$ are the mass-loss and mass-gain rates, respectively, and $\psi$ is the SFR. 
Two additional assumptions are made. First, we assume that the accreted gas is metal-free and the outflowing gas has the same metallicity as the ISM at the time the outflows are launched. Second, we require the mass loading and mass accretion factors to be constant. 
Note that this assumption only constraints the factors (ratios) to be constant, and does not restrict the time evolution of star formation, inflow rate and outflow rate to any specific forms. 

The first assumption of outflows is valid if most of the outflowing gas is entrained ISM close to the energy sources (e.g. supernovae). Indeed, more than 75\% of the outflowing gas is estimated to be entrained gas in nearby mergers \citep{Rupke:2013zr}. The second assumption of constant mass-loading and mass accretion factors helps the models to remain analytic and simple, and the assumption carries significant physical meanings. The constant mass-loading factor is postulated because the energy driving the outflowing mass is from star formation, and $\eta$ simply reflects the efficiency in transferring energy from star formation to the outflowing gas. Similarly, because the inflowing gas supplies the reservoir for star formation, the constant mass accretion factor can be realised as the efficiency of collapsing the gas into stars. We explore different $\eta$ and $\Lambda$ values later to understand the possible impact of them not being constant over time. \citet{Recchi:2008ly} explore non-constant $\Lambda$ by assuming exponential inflows and a linear Schmidt law. They conclude that constant $\Lambda$ is a reasonable approximation of the late evolution of a galaxy provided that the infall timescale is of the same order of magnitude of the star formation timescale.

Under these assumptions, a pair of non-negative $(\eta,\Lambda)$ determines a unique analytical solution for the metallicity and the stellar-to-gas mass ratio:
\begin{equation}\label{main-eq-solution}
Z(t)={y \over \Lambda}\left\{1-\left[1+ \left(1+{\eta-\Lambda\over1-R}\right) {M_{*o}(t)\over M_g(t)}\right]^ {-{\Lambda \over 1-R + \eta-\Lambda}}\right\} 
\end{equation}
where
\begin{equation}\label{main-eq-conditions}
\eta \geq 0, \ \Lambda > 0, \ {\rm and }\ {\eta-\Lambda\over1-R} \neq -1. 
\end{equation}
That is, the stellar-to-gas mass ratio at any given time dictates the metallicity at that instant. The reader is referred to Kudritzki et al. (in preparation) for derivation of the models and special cases when the conditions in Equation~\ref{main-eq-conditions} are not met. We also show in Kudritzki et al. (in preparation) that in the trivial case where there are no inflows nor outflows, i.e. $\eta=0$ and $\Lambda =0$, our model is identical to the classical closed-box model (i.e. Equation~\ref{main-eq-closed-box}).

\begin{figure*}
\centering
\includegraphics[width=16cm]{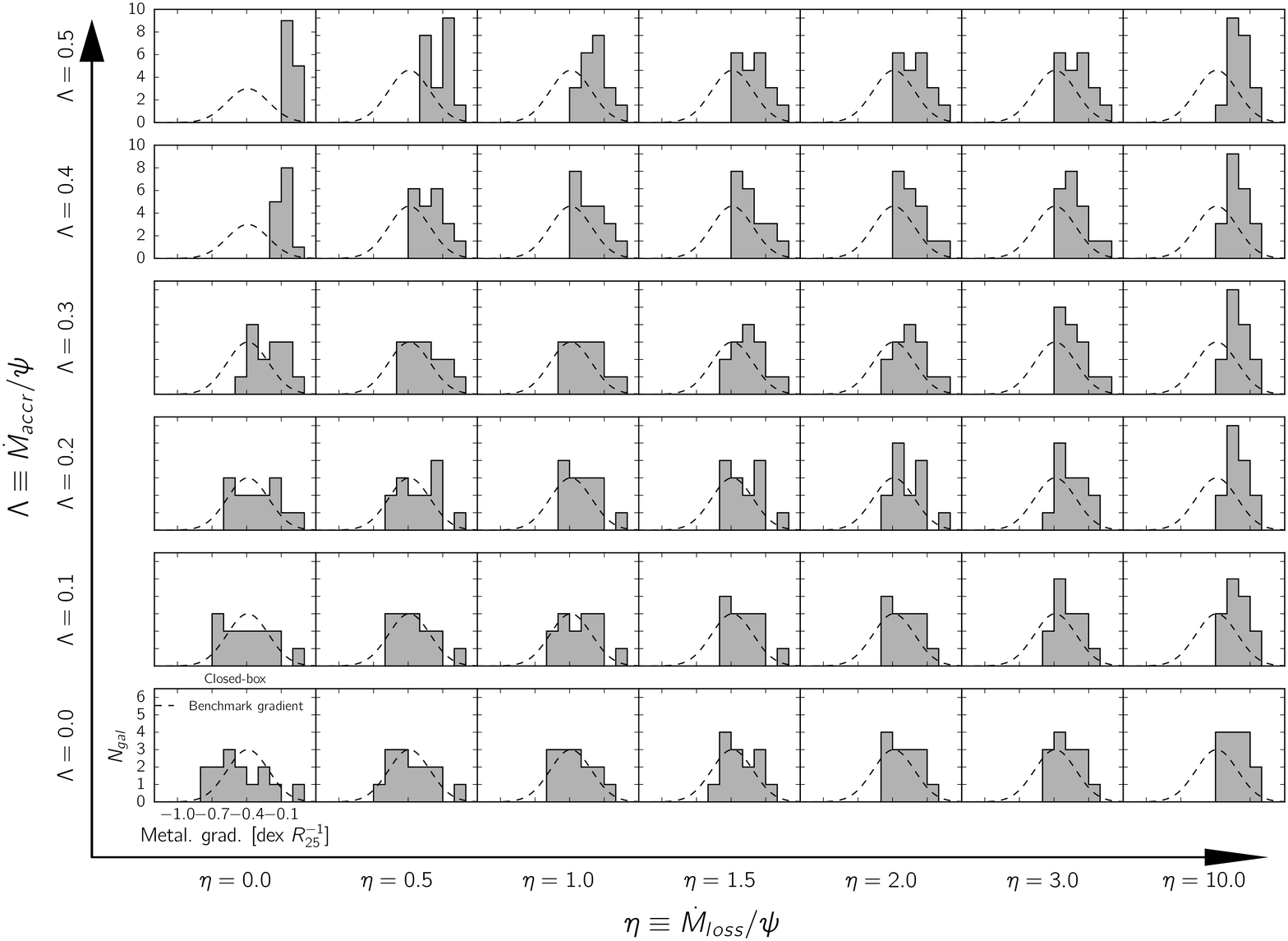}
\caption{ Distributions of the predicted metallicity gradients of the 14 field spiral galaxies shown in Figure~\ref{ms_mg_profile}. The metallicity gradients are predicted using the analytical models described in the text and in Kudritzki et al. (in preparation). A grid of mass loading factor $\eta$ and mass accretion factor $\Lambda$ is adopted to predict the metallicity gradients. The $\eta$ and $\Lambda$ values are labeled in the outer, large axes. The closed-box model is at  $(\eta,\Lambda)=(0,0)$. The measured benchmark gradient is shown as the black dashed curves for comparison. All the panels, except for the two top-left panels, have the same scales, as indicated in the bottom-left panel; the y-scales of the two top-left panels are labeled separately to accommodate the concentration of flat metallicity gradients in one bin. }\label{predict_gradients}
\end{figure*}

In Figure~\ref{modelExamples}, we present the relationships between the metallicity and stellar-to-gas mass ratio (lower panel) using different sets of $(\eta,\Lambda)$ values (upper panel). We adopt the yield for the oxygen of 0.00313 and the returned mass fraction of 0.4, and we will discuss the systematics of these two constants later. The classical closed-box model, which corresponds to the origin on the $\eta\ \mbox{-}\ \Lambda$ plane (upper panel), is also shown as the dashed line in the lower panel. In the gas rich regime, i.e. ${\log}(M_{*o}/M_{g})\ll-0.5$, all the models coalesce as, under vast gas reservoirs, inflows and outflows do not change the metallicity appreciably. In the gas poor regime, i.e. ${\log}(M_{*o}/M_{g})\gg-0.5$, for a given stellar-to-gas mass ratio the metallicity is sensitive to the adopted mass loading and mass accretion factors. This regime can be explored with the the 14 field spiral galaxies from \citet{Leroy:2008qy} ($0\lesssim{\log}(\Sigma_{*}/\Sigma_{g}) \lesssim2$; c.f. Figure~\ref{ms_mg_profile}), which could place constraints on the mass loading and mass accretion factors.

By combining the models and the measured stellar-to-gas mass profiles of the 14 field spiral galaxies, we can predict their metallicity gradients. The predicted metallicity gradients can be compared with our benchmark gradient to place constraints on the models. We calculate the predicted metallicity gradients by first converting the stellar-to-gas mass ratios to metallicities for each radial bins, and we fit linear profiles to data at $r>0.2R_{25}$ in each galaxies to derive the predicted metallicity gradients. In Figure~\ref{predict_gradients}, we show the distributions of the predicted metallicity gradients using a grid of ($\eta, \Lambda$). In the top and middle panel of Figure~\ref{diffMeanSigma}, we compare the means and standard deviations of the predicted metallicity gradients to those from our benchmark gradient. In the bottom panel, we perform one-sample Kolmogorov-Smirnov tests to compare the distributions of the predicted metallicity gradients with the benchmark gradient that has a Gaussian distribution of $-0.39\pm0.18~{\rm dex}~R_{25}^{-1}$ (mean $\pm$ standard deviation; Figure~\ref{gradient-benchmark}).

Figure~\ref{predict_gradients} qualitatively demonstrates that small $\eta$ and $\Lambda$ values (panels toward the lower left corner) are preferred because the predicted distributions are similar to the benchmark gradient. Large $\eta$ and $\Lambda$ values tend to overproduce flatter metallicity gradients, effectively shifting the means of the distributions towards zero and reducing the widths of the distributions. This behaviour can be trivially understood with the bottom panel of Figure~\ref{modelExamples} where the models with large $\eta$ and $\Lambda$ values flatten at high stellar-to-gas mass ratios, yielding the same metallicities across the disks, i.e. flat metallicity gradients. We quantitatively address the allowed $\eta$ and $\Lambda$ values in Figure~\ref{diffMeanSigma} by investigating the differences of the means (upper panel), those of the standard deviations (middle panel), and the probabilities of reproducing the benchmark gradient through the Kolmogorov-Smirnov test (bottom panel). We find that for $0\lesssim\Lambda\lesssim0.2$ and $0\lesssim\eta\lesssim2$, the Kolmogorov-Smirnov tests yield good probabilities ($\gtrsim20\%$) for the distributions of the predicted metallicity gradients to be drawn from the benchmark gradient. Within the same ranges, the differences between the mean of the benchmark gradient and those of the predicted gradients are within about $20\%$, and the differences between the standard deviations are also within about $20\%$. Interestingly, the differences of the means, and those of the standard deviations both show that the closed-box model is the best model, but the Kolmogorov-Smirnov tests suggest that low (but non-zero) mass loading and mass accretion factors are more preferred. While the precise values of $\eta$ and $\Lambda$ probably cannot be determined from the 14 galaxies alone (due to low number statistics) and are likely to vary from system to system, closed-box and virtually closed-box are the models that can successfully reproduce the observed benchmark gradient.

The success of reproducing the benchmark gradient with our simple models, however, do not imply that all galaxies evolve as closed-box or virtually closed-box throughout their lifetime. Nor do our results support the idea that galaxies always have constant mass loading and mass accretion factors, either on global or spatially resolved scales. Observations of high-redshift galaxies ($z>1$) have provided evidences that galaxies in the early Universe undergo many, perhaps intermittent, accretion events, immense star formation and outflows \citep[e.g.,][]{Tacconi:2010fk,Tacconi:2013ly,Weiner:2009lr,Steidel:2010fk,Genzel:2011fk,Newman:2012nx,Genzel:2014fk}. Similar outflows, in particular the starburst-driven winds, are found to be ubiquitously in galaxies at lower redshifts with high enough star-formation surface density \citep{Heckman:2002fk}, with the wind velocities showing indication of correlating with both the SFR and host galaxy mass \citep{Rupke:2005fv,rupke:2005b,Veilleux:2005qy,Chen:2010qy}. Energy and mass return from both the stars and AGNs (i.e. ``feedback'') are indispensable for numerical simulations to reproduce many observed properties of galaxies, such as the stellar mass function and mass-metallicity relation \citep[e.g.,][]{Springel:2003dq,Oppenheimer:2010lr,Dave:2011qy,Dave:2011fk}. From the theoretical considerations, \citet{Murray:2005fk} suggest that the mass loading factor could vary with the host galaxy mass, following different scaling relations depending on the winds being energy or momentum-driven. The later is favoured by recent smoothed-particle hydrodynamics + N-body simulations on both galactic and cosmological scales \citep[e.g.,][]{Dave:2011qy,Dave:2011fk,Hopkins:2012uq}, but the two mechanisms dominating in galaxies of different masses has also been suggested \citep{Dutton:2009rr}. Unfortunately, measuring the mass loading factor accurately from observations remains difficult and a consensus on its values has not been reached yet \citep{Zahid:2014rt}.  The multi-phase nature of the outflowing gas that spans wide ranges in both density and temperature poses a major observational challenge (see \citealt{Veilleux:2005qy} for a review).

\begin{figure}
\centering
\includegraphics[width=8.6cm]{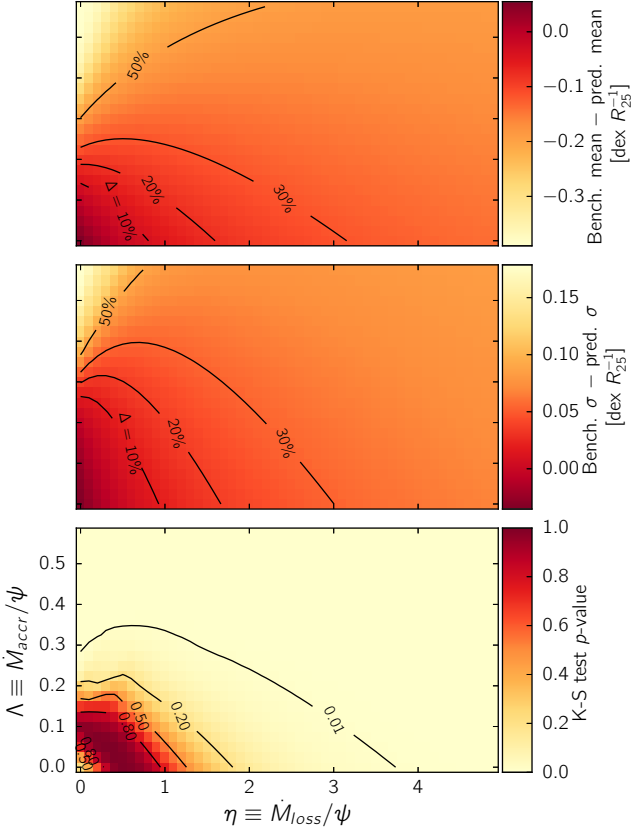}
\caption{
A quantitative comparison between the benchmark gradient and the metallicity gradients predicted using the 14 field spiral galaxies from \citet{Leroy:2008qy} and our analytical models. A qualitative comparison is also presented in \ref{predict_gradients}. Each location on the plots corresponds to adopting one set of $(\eta,\Lambda)$ values to predict the 14 metallicity gradients. The top panel shows the differences in the means, and the middle panel shows the differences in the standard deviation, with contours indicating the differences in percentage. The bottom panel shows the {\it p-}values from one-sample Kolmogorov-Smirnov tests. A higher {\it p-}value indicates a higher probability for the distribution of the 14 predicted metallicity gradients to be drawn from the benchmark gradient that has a Gaussian distribution of $-0.39\pm0.18~{\rm dex}~R_{25}^{-1}$ (mean $\pm$ standard deviation). }\label{diffMeanSigma}
\end{figure}

Despite the complexity and the lack of observational constraints on outflows and inflows, our simple models still can reproduce the benchmark gradient because the metallicity is not sensitive to the adopted mass loading and accretion factors in the gas rich regime (${\log}(M_{*o}/M_{g})\ll-0.5$; Figure~\ref{modelExamples}). Recent radio observations of $z>1$ galaxies reveal that galaxies at high redshifts are typically gas rich \citep{Tacconi:2010fk,Tacconi:2013ly}, and therefore the potentially high mass loading and mass accretion factors at high redshifts do not determine the metallicity gradients at $z=0$. We hypothesise that as galaxies evolve to higher stellar-to-gas ratio (presumably at $z <1$), both the mass loading and accretion factors decrease dramatically and stabilise such that their chemical evolution can be approximated by the closed-box or virtually closed-box models. 

Our analysis favours a very low mass accretion factor ($\Lambda\lesssim0.3$), consistent with the lack of direct observational evidence of gas accretion in field galaxies in the local Universe. The low mass accretion factor and high stellar-to-gas mass ratio imply that field star-forming galaxies in the local Universe have no significant, recent refuelling of their gas reservoirs and their low level of star formation activities are sustained by the remaining gas reservoirs acquired presumably at high redshift. We also obtain a marginally low mass loading factor of about $\eta\lesssim2$. Such mass loading factor is consistent with the range of $\eta$ measured. \citet{Zahid:2012ys} empirically constrain the mass loading factor in star-forming galaxies to be less than 1 by assuming that these galaxies evolve on the measured mass-metallicity and the galaxy main-sequence \citep{Noeske:2007lr}. \cite{Bolatto:2013qy} estimate the mass loading factor (of the molecular gas) of more than 1 (and probably $\sim3$) in the nearby starburst galaxies NGC~253. Mass loading factors of about 0.1--1 were also found in $z<0.5$ luminous and ultra-luminous infrared galaxies, and nearby mergers \citep[][]{rupke:2005b,Rupke:2013zr}.

We note that although we assumed the oxygen yield as constant, the oxygen yield varies with both the stellar metallicity and initial mass function (e.g., \citealt{Maeder:1992bh,Woosley:1995lq,Kobayashi:2006qy}; see \citealt{Zahid:2012ys} for a summary). We also assumed a constant returned mass fraction, but the returned mass fraction is  functions of both the stellar age and initial mass function, spanning a range of approximately $0.15\mbox{--}0.45$ \citep[e.g.,][]{Leitner:2011lr}, and $0.3\mbox{--}0.45$ for the Salpeter and Chabrier initial mass function \citep{Salpeter:1955kx,Chabrier:2003uq}. Constraining these two parameters individually has proven to be difficult as they are degenerate through $y/(1-R)$, i.e. the pre-factor in Equation~\ref{main-eq-closed-box} (see also \citealt{Zahid:2012ys}). A higher degree of nucleosynthesis of the oxygen  (higher $y$) can be balanced by locking up more oxygen in each generation of stars (higher $1-R$; lower $R$), effectively leaving the same amount of oxygen in the ISM. In this work, we adopt the oxygen yield and return gas fraction from Kudritzki et al. (in preparation) who empirically constrain $y/(1-R)$ to the accuracy of 25\% by reproducing the metallicity and the metallicity gradient of the young stellar population in the Milky Way. We vary the returned mass fraction from 0.15 to 0.45 while keeping $y/(1-R)$ fixed, which corresponds to an oxygen yield between 0.0045 to 0.0029, and our results do not change considerably. We find that a higher (lower) returned mass fraction resulting in more (less) gas return would flatten (steepen) the model curves in the bottom panel of Figure~\ref{modelExamples} at high stellar-to-gas mass ratio. However, the degree of flattening (steepening) is insignificant such that similar preferred $\eta$ and $\Lambda$ are recovered. For reasonable returned mass fractions ($R=[0.45,0.3,0.15]$), our preferred $\eta$ and $\Lambda$ ranges, defined by the 20\% contour of the Kolmogorov-Smirnov tests, remain virtually the same ($0\lesssim\Lambda\lesssim[0.2,0.25,0.3]$; $0\lesssim\eta\lesssim[1.8,2.1,2.8]$).

In the models, we assumed that the outflowing gas has the same metallicity as the ISM at the time the outflows are launched. This assumption is appropriate because more than 75\% of the outflowing mass in nearby mergers is entrained gas \citep{Rupke:2013zr}. Evidences of the hot, wind fluid being more enriched than the ISM have been reported at least in one nearby dwarf galaxy NGC~1569 \citep{Martin:2002lr}, perhaps indicating that the hot materials can survive the gravitational potential better than the cold entrained gas. Constraints on the metallicity of the outflowing gas remain scarce because X-ray observations are often required. If indeed the outflowing gas has a higher metallicity than the ISM, the metallicity at a given stellar-to-mass ratio would be lower than that without a higher metallicity, particularly when the stellar-to-mass ratio is high. More enriched galactic winds, similarly, will flatten the model curves in the bottom panel of Figure~\ref{modelExamples}, causing our analysis to favour an even smaller mass loading factor.

\section{Summary and Conclusions}\label{sec-summary}

We have presented metallicity gradients of 49 local field star-forming galaxies measured with integral field spectroscopy and slit spectroscopy. Metallicities have been determined for these galaxies using strong optical emission lines ({[\ion{O}{ii}]~$\lambda\lambda$3726,3729}, H$\beta$, {[\ion{O}{iii}]~$\lambda$5007}, H$\alpha$, and {[\ion{N}{ii}]~$\lambda$6583}) with two widely adopted metallicity calibrations (the O3N2 diagnostic by \citealt{Pettini:2004lr}; and the N2O2 diagnostic by \citealt{Kewley:2002fj}). Our results show that the metallicities measured with the two calibrations are typically in good agreement ($\pm0.1$~dex), but the differences in metallicities correlate with the ionisation parameters. Similarly, the two calibrations yields metallicity gradients typically in good agreement ($\pm0.14~{\rm dex}~R_{25}^{-1}$), but up to $0.4~{\rm dex}~R_{25}^{-1}$ difference is possible when the ionisation parameters change systematically with radius. 

When comparing the metallicity gradients with the stellar masses and absolute {\it B}-band magnitudes, we find that, when the metallicity gradients are expressed in $\rm dex~kpc^{-1}$, galaxies with lower masses and luminosities have (1) on average a steeper metallicity gradient and (2) more diverse metallicity gradients compared to galaxies of higher masses and luminosities. Such dependencies on mass and luminosity do not exist when the sizes of galaxies are taken into account and the metallicity gradients are expressed in terms of ${\rm dex}~R_{25}^{-1}$. All our disk galaxies appear to have a common metallicity gradient when normalised to the optical radii of the galaxies, consistent with previous studies. This leads us to quantify a local benchmark gradient of $-0.39\pm0.18~{\rm dex}~R_{25}^{-1}$ that could be useful for comparison with metallicity gradients measured at high redshifts. 

We adopt simple chemical evolution models to investigate the cause of the common, uniform metallicity gradients. Starting from the measured atomic and molecular gas, and stellar surface density profiles in 14 nearby, field spiral galaxies, our analytical models can qualitatively and quantitatively reproduce the measured local benchmark gradient. 
Our results suggest that the galactic disks of spiral galaxies (at $0.2\lesssim r / R_{25}\lesssim1$) evolve chemically close to the closed-box model when the stellar-to-gas ratio becomes high (${\log}(M_{*o}/M_{g})\gg-0.5$). The inferred negligible mass accretion rates ($\lesssim0.3 \times\rm SFR$), and very low mass outflow rates ($\lesssim3\times\rm SFR$) are broadly consistent with observational constraints.

To summarise, our simple chemical models already capture the fundamental physics governing the common metallicity gradient. The common metallicity gradient is a direct result of the common gas and stellar surface density profiles under the coevolution of gas, stars, and metals during galaxies build up their mass.

\section*{Acknowledgments}

We thank the anonymous referee for constructive comments and suggestions. MAD and LJK acknowledge the support of the Australian Research Council (ARC) through Discovery project DP130103925. MAD also acknowledges financial support from King Abdulaziz University under the HiCi program. RPK and FB were supported by the National Science Foundation under grant AST-1008798. We thank support from the Time Assignment Committee at the Research School of Astronomy and Astrophysics of the Australian National University.

This study makes uses of the data provided by the Calar Alto Legacy Integral Field Area survey (\url{http://califa.caha.es/}). Based on observations collected at the Centro Astron\'{o}mico Hispano Alem\'{a}n (CAHA) at Calar Alto, operated jointly by the Max-Planck-Institut f\"{u}r Astronomie and the Instituto de Astrofisica de Andalucia (CSIC). This research used NASA's Astrophysics Data System Bibliographic Services and the NASA/IPAC Extragalactic Database (NED). We acknowledge the usage of the HyperLeda database (\url{http://leda.univ-lyon1.fr}).

\appendix

\section{Metallicity gradients and ionisation parameters of Individual galaxies}\label{appendix}
Figures~\ref{more_califa} and \ref{more_s12} show the metallicity gradients and ionisation parameters for the rest of the CALIFA and S12 samples, respectively. These are in addition to the four galaxies presented in Figures~\ref{califa_gradient} and  \ref{s12_gradient}. Figure~\ref{wifes_gradient} show those for the WiFeS galaxies. 

\begin{figure*}
\centering
\includegraphics[width=16.5cm]{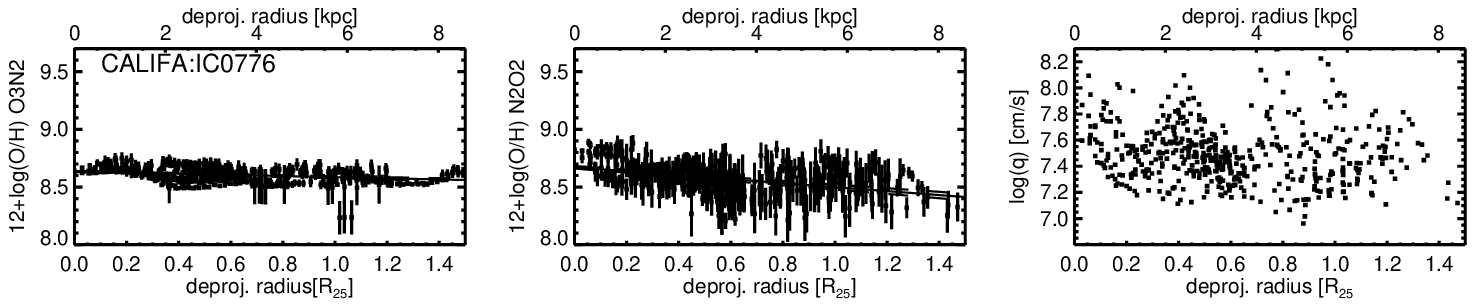}
\includegraphics[width=16.5cm]{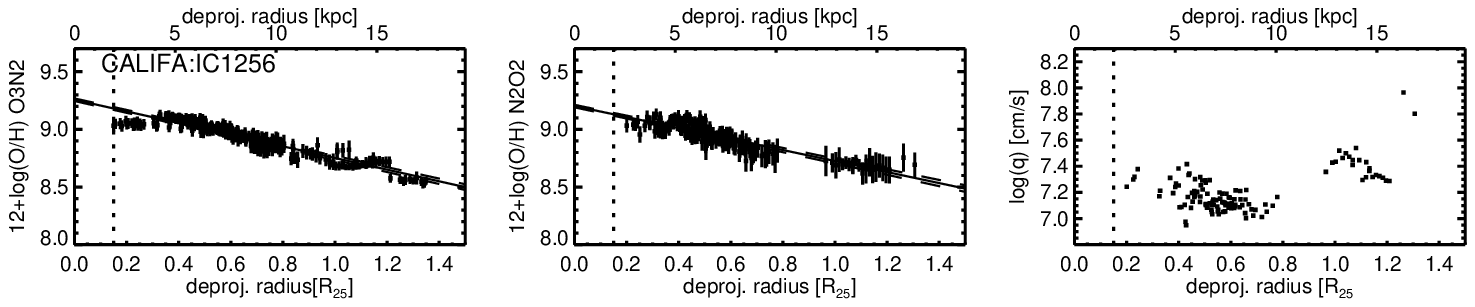}
\includegraphics[width=16.5cm]{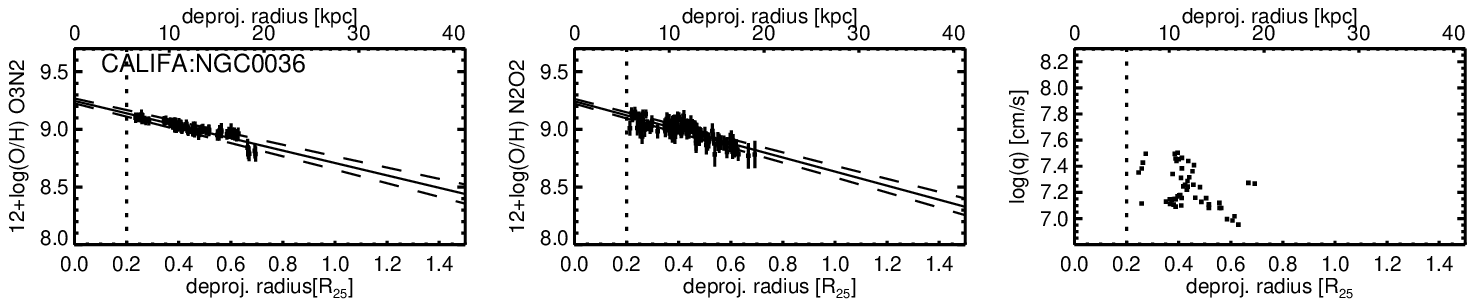}
\includegraphics[width=16.5cm]{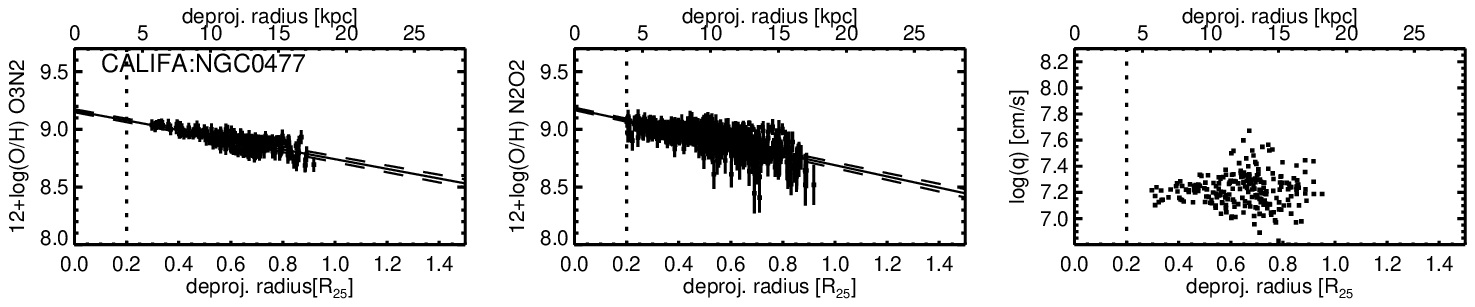}
\includegraphics[width=16.5cm]{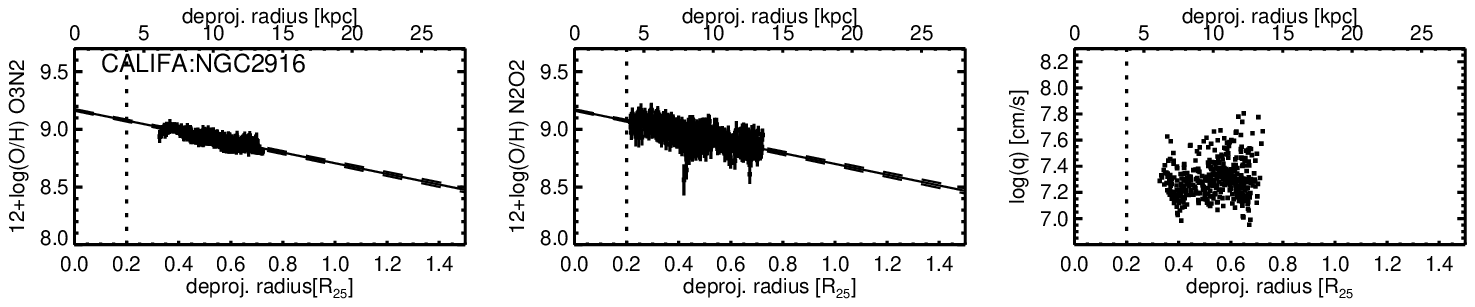}
\includegraphics[width=16.5cm]{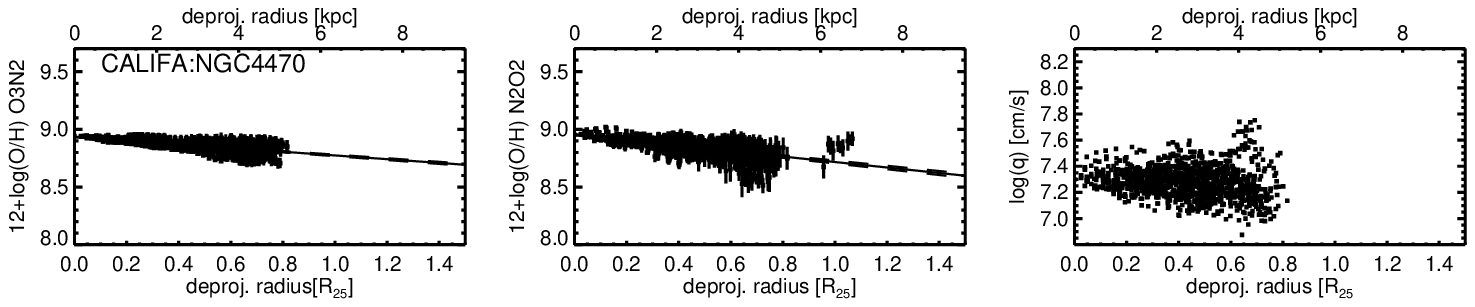}
\includegraphics[width=16.5cm]{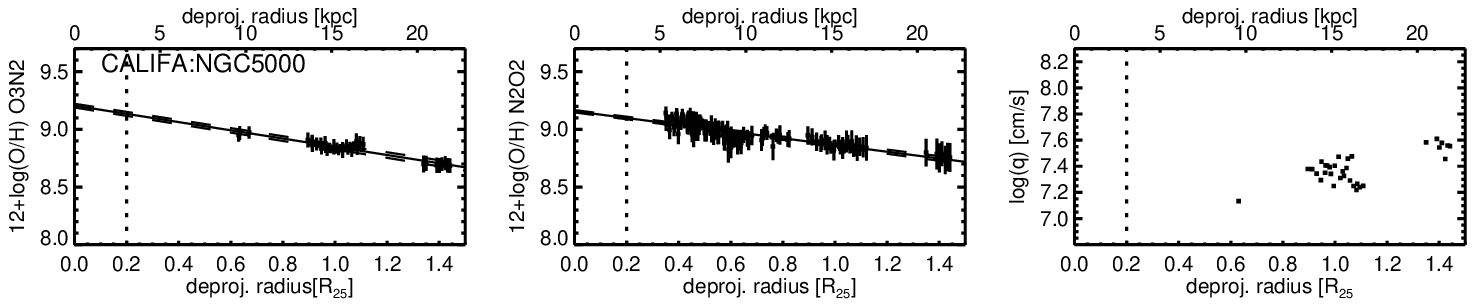}

\caption{Continuation of Figure~\ref{califa_gradient}. }\label{more_califa}
\end{figure*}
\addtocounter{figure}{-1}

\begin{figure*}
\centering
\includegraphics[width=16.5cm]{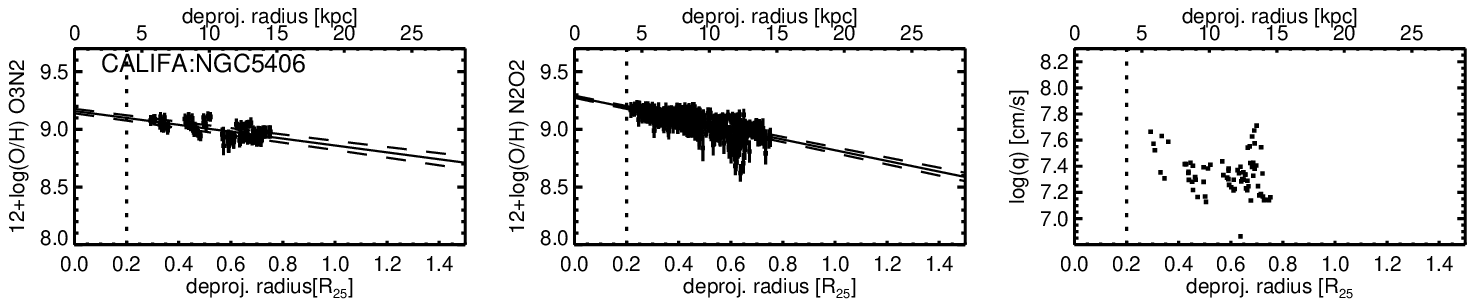}
\includegraphics[width=16.5cm]{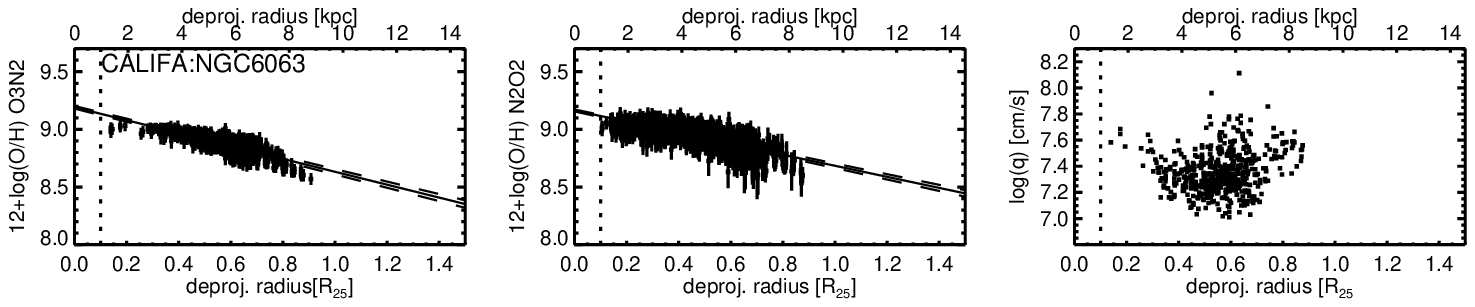}
\includegraphics[width=16.5cm]{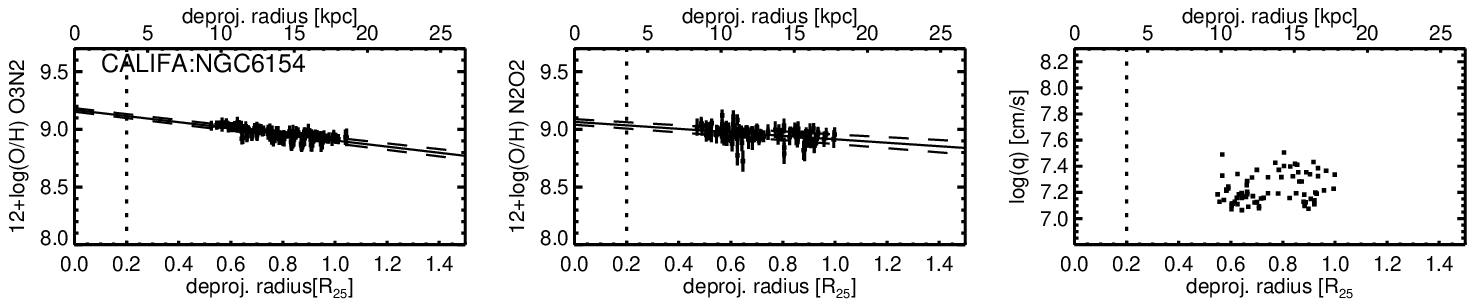}
\includegraphics[width=16.5cm]{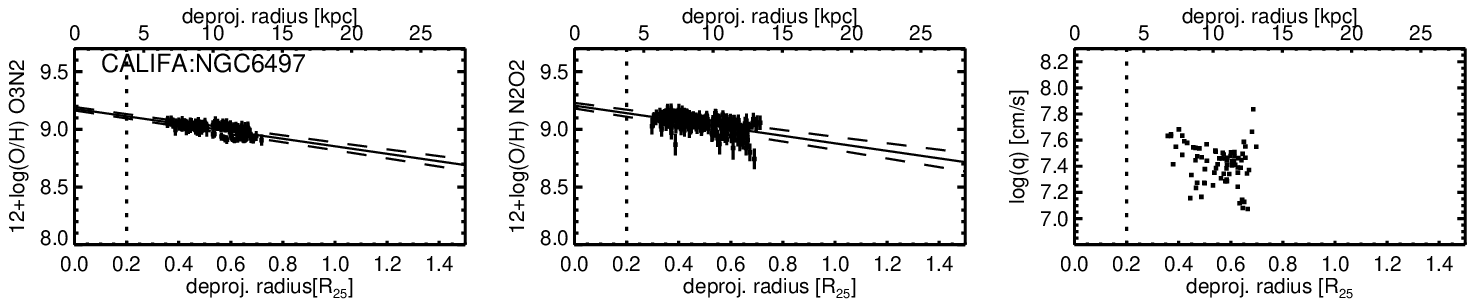}
\includegraphics[width=16.5cm]{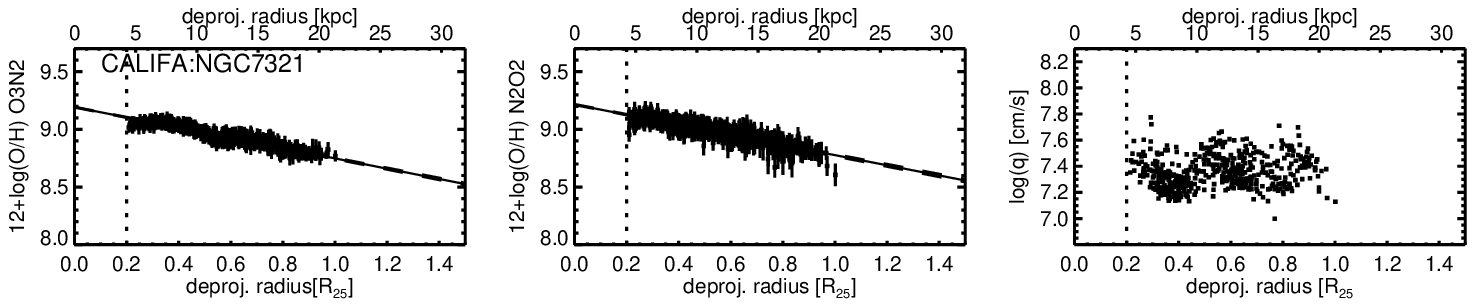}
\includegraphics[width=16.5cm]{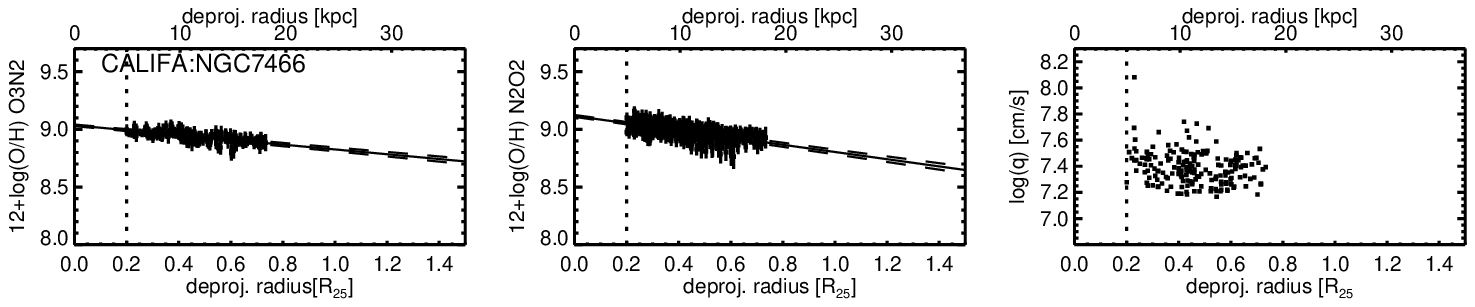}
\includegraphics[width=16.5cm]{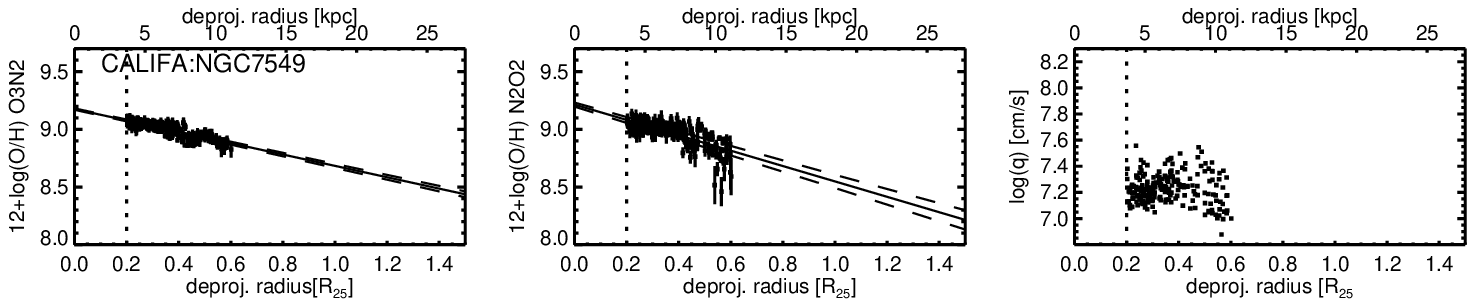}
\caption{{\it Continue}}
\end{figure*}
\addtocounter{figure}{-1}

\begin{figure*}
\centering
\includegraphics[width=16.5cm]{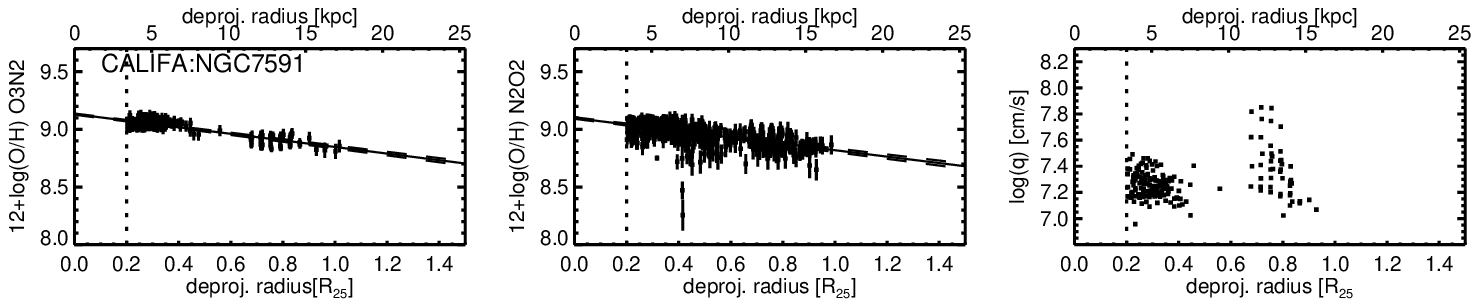}
\includegraphics[width=16.5cm]{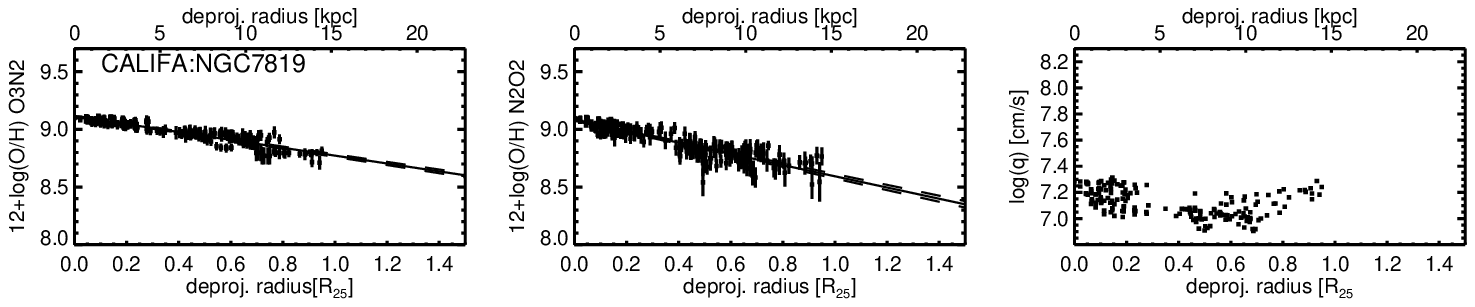}
\includegraphics[width=16.5cm]{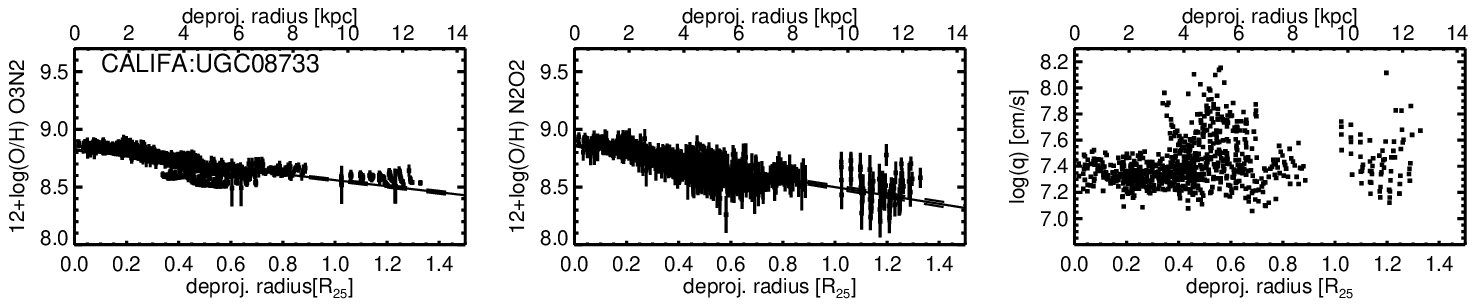}
\includegraphics[width=16.5cm]{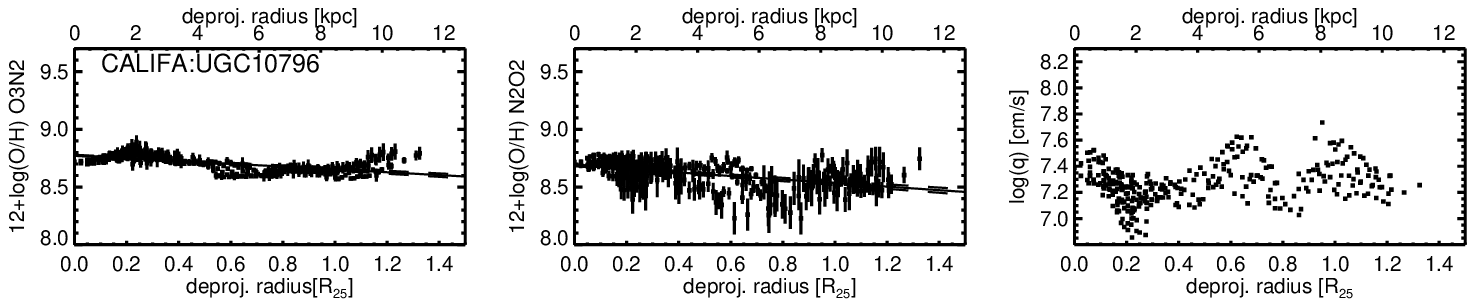}
\includegraphics[width=16.5cm]{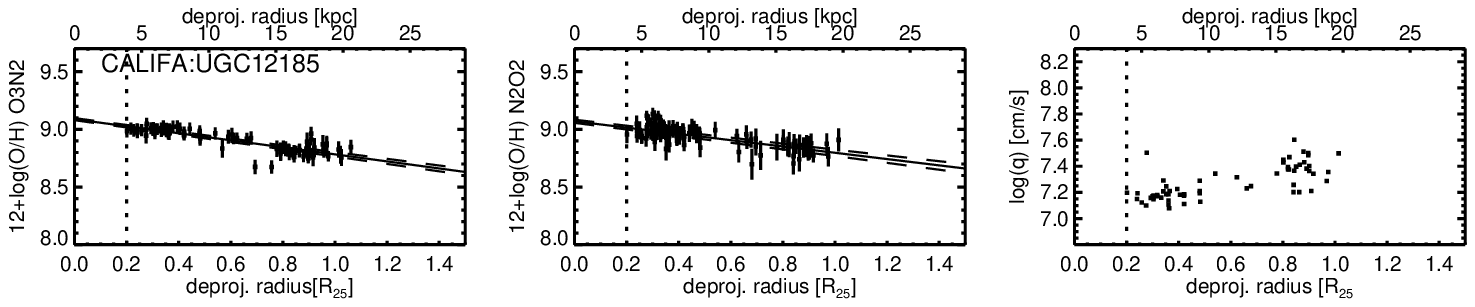}
\caption{{\it Continue}}
\end{figure*}

\begin{figure*}
\centering
\includegraphics[width=16.5cm]{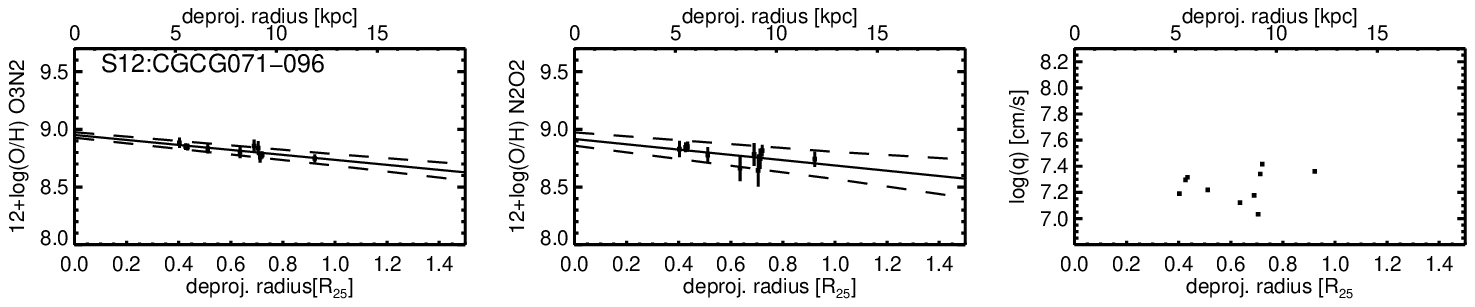}
\includegraphics[width=16.5cm]{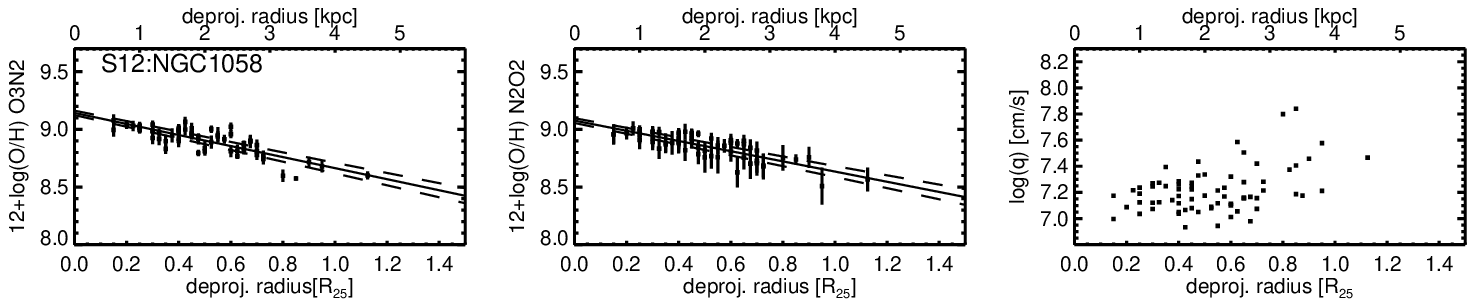}
\includegraphics[width=16.5cm]{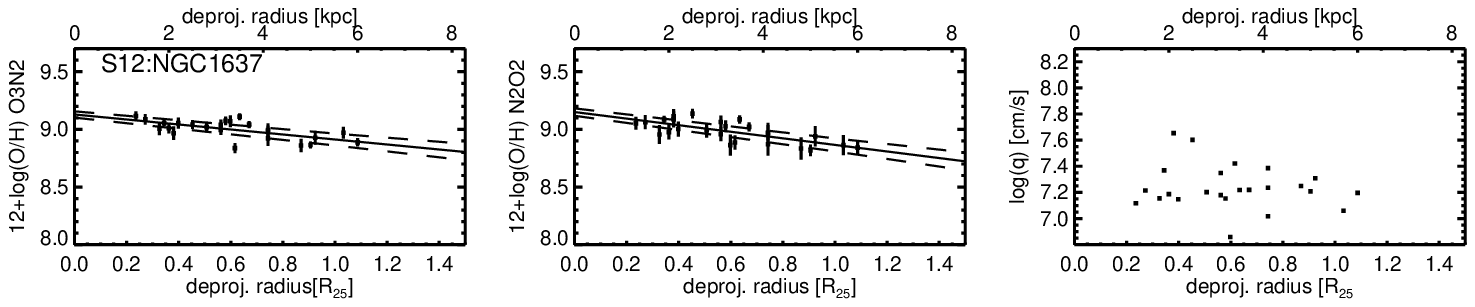}
\includegraphics[width=16.5cm]{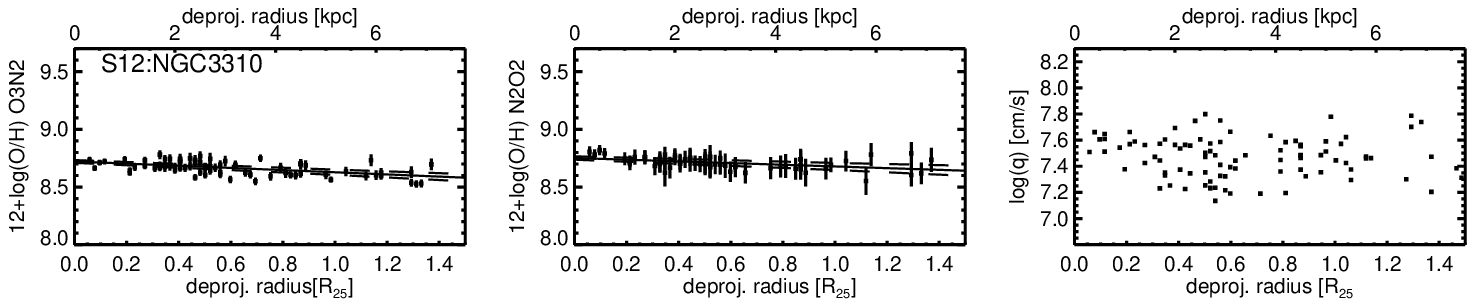}
\includegraphics[width=16.5cm]{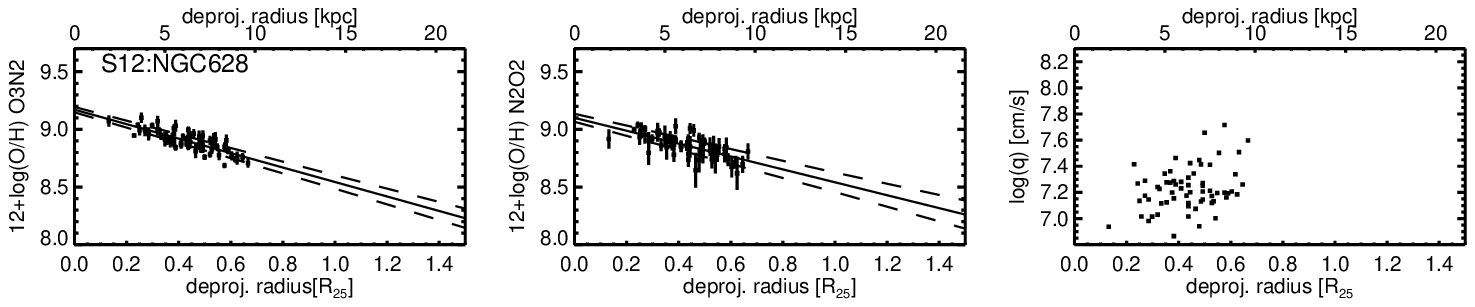}
\includegraphics[width=16.5cm]{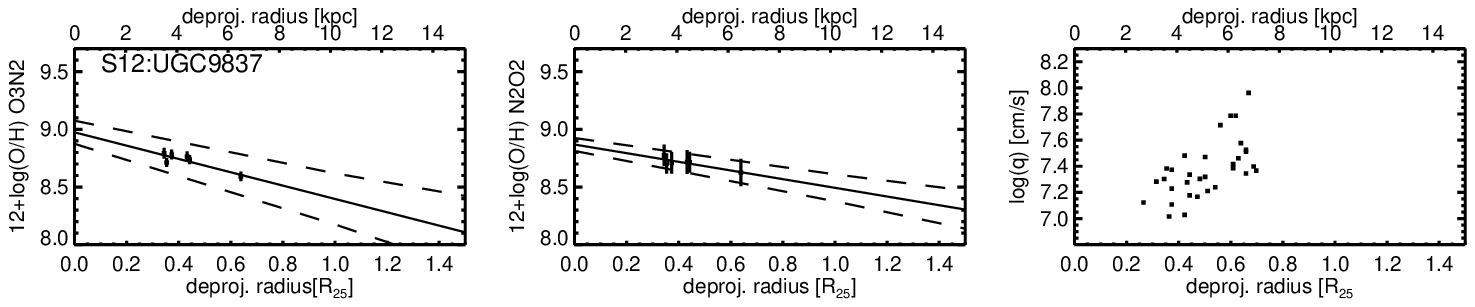}
\includegraphics[width=16.5cm]{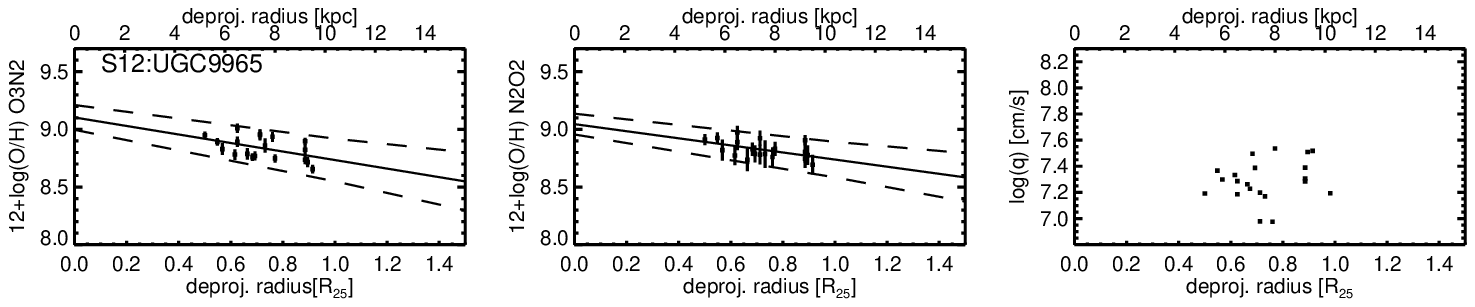}
\caption{Continuation of Figure~\ref{s12_gradient}. }\label{more_s12}
\end{figure*}

\begin{figure*}
\centering
\subfloat{\includegraphics[width = 8cm]{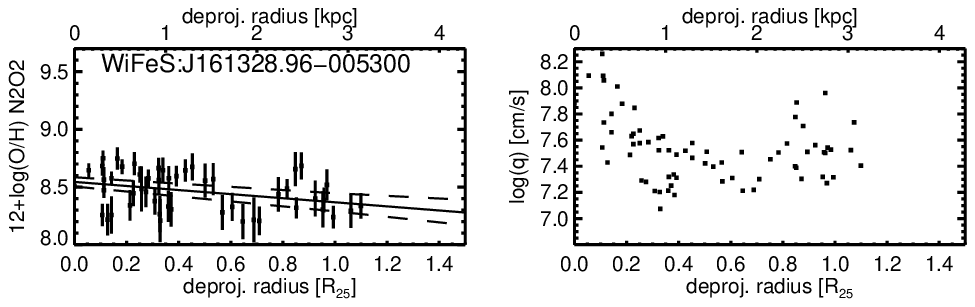}} \subfloat{\includegraphics[width = 8cm]{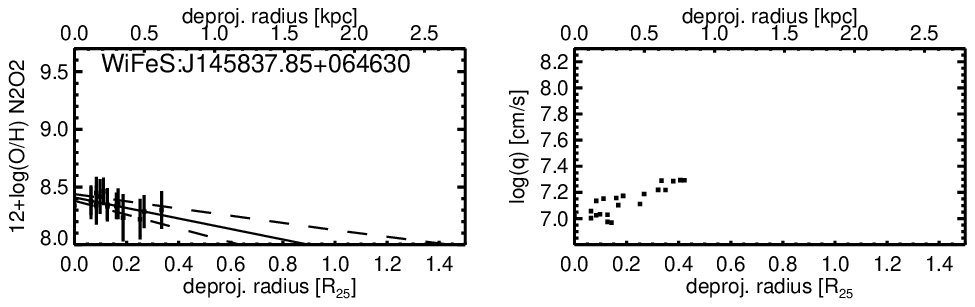}}\\

\subfloat{\includegraphics[width = 8cm]{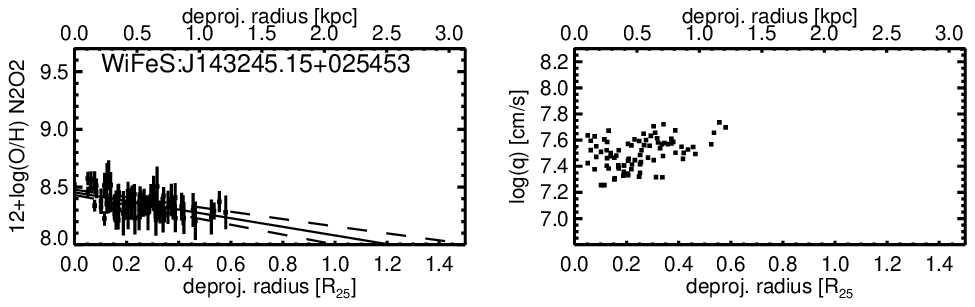}} \subfloat{\includegraphics[width = 8cm]{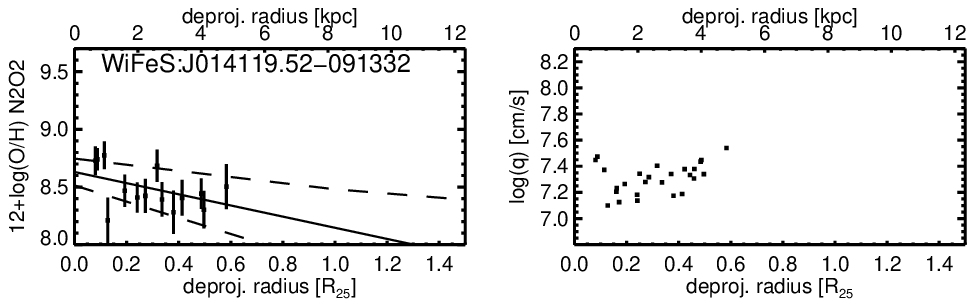}}\\

\subfloat{\includegraphics[width = 8cm]{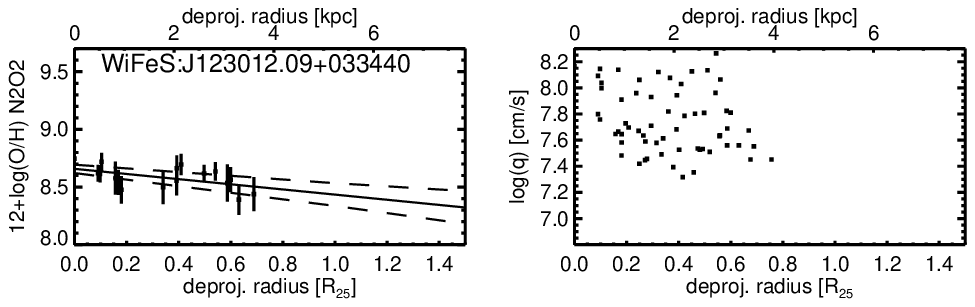}} \subfloat{\includegraphics[width = 8cm]{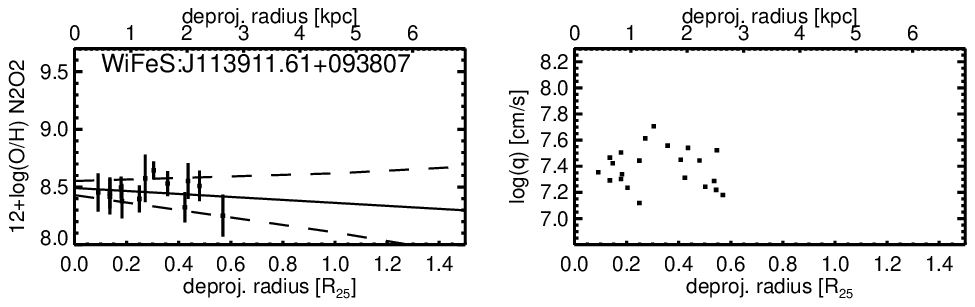}}\\

\subfloat{\includegraphics[width = 8cm]{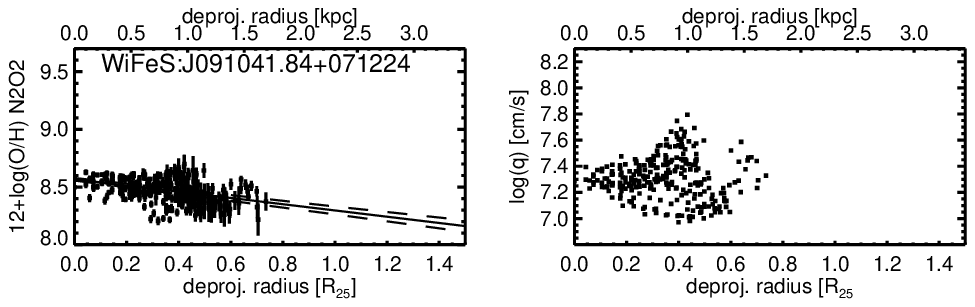}} \subfloat{\includegraphics[width = 8cm]{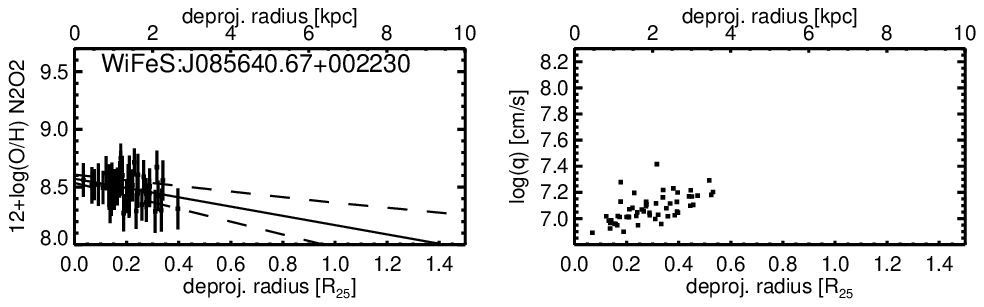}}\\

\subfloat{\includegraphics[width = 8cm]{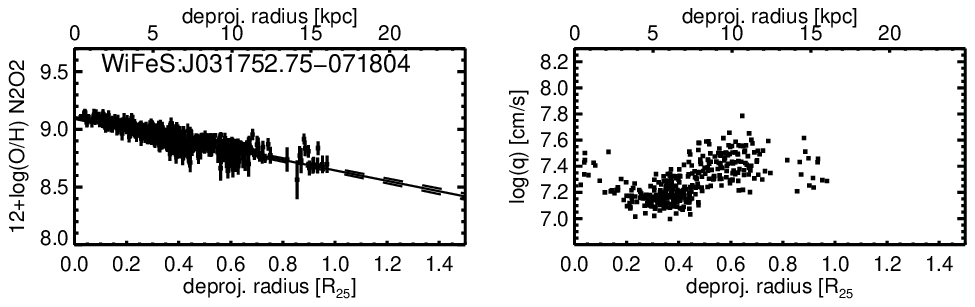}} \subfloat{\includegraphics[width = 8cm]{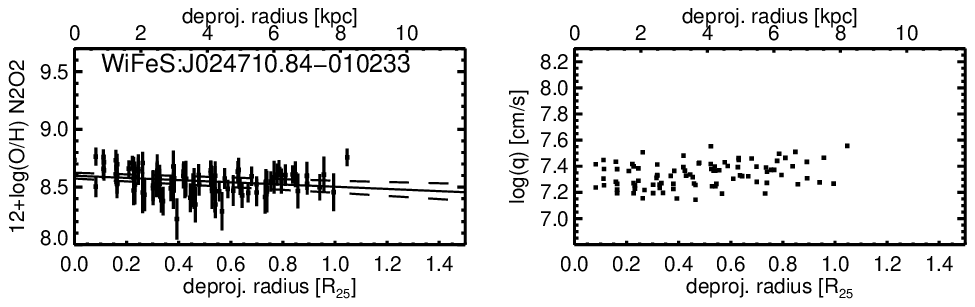}}\\

\caption{Same as Figure~\ref{califa_gradient}, but for the WiFeS galaxies. Only the N2O2 metallicity gradients (left panels) and ionisation parameter versus radius (right panels) are shown. See more details in Section~\ref{sec-metallicity}. }\label{wifes_gradient}
\end{figure*}

\bibliography{/Users/itho/Dropbox/references}

\end{document}